\documentclass{aa}  
\usepackage{graphicx}
\usepackage{soul}
\usepackage{placeins}
\usepackage{txfonts}
\usepackage{xcolor}
\usepackage{amsmath,amssymb, graphicx, multicol, array, upgreek, natbib, booktabs, multirow, bm, dsfont}

\usepackage{graphicx}
\usepackage{wrapfig}

\usepackage{tabularx}
\setstcolor{red}

\begin{document}

\title{Cosmology from LOFAR Two-metre Sky Survey Data Release 2:
Cross-correlations with luminous red galaxies from eBOSS}
\titlerunning{Cosmology from LOFAR LoTSS DR2: Cross-correlation with LRGs from eBOSS}
\authorrunning{J.~Zheng et al.}

\author{Jinglan Zheng\inst{1}\thanks{jzheng@physik.uni-bielefeld.de}
\and
Prabhakar Tiwari\inst{2}\thanks{prabhakar.tiwari@gtiit.edu.cn}
\and
Gong-Bo Zhao\inst{3,4,5}\thanks{gbzhao@nao.cas.cn}
\and
Dominik J.~Schwarz\inst{1}
\and
David Bacon\inst{6}
\and
Stefano Camera\inst{7,8,9}
\and
Caroline Heneka\inst{10}
\and
Catherine Hale\inst{11}
\and
Szymon J.~Nakoneczny\inst{12}
\and
Morteza Pashapour-Ahmadabadi \inst{1}
}

\institute{Fakult\"at f\"ur Physik, Universit\"at Bielefeld, Postfach 100131, 33501 Bielefeld, Germany
\and 
Department of Physics, Guangdong Technion - Israel Institute of Technology, Shantou, Guangdong 515063, P.R.\ China
\and
National Astronomy Observatories, Chinese Academy of Sciences, Beijing, 100101, P.R.\ China
\and 
School of Astronomy and Space Science, University of Chinese Academy of Sciences, Beijing 100049, P.R.\ China
\and 
Institute for Frontiers in Astronomy and Astrophysics, Beijing Normal University, Beijing 102206, P.R.\ China
\and
Institute of Cosmology and Gravitation, University of Portsmouth, Burnaby Road, Portsmouth PO1 3FX, United Kingdom
\and
Dipartimento di Fisica, Universit\`a degli Studi di Torino, Via P.\ Giuria 1, 10125 Torino, Italy
\and
INFN -- Istituto Nazionale di Fisica Nucleare, Sezione di Torino, Via P.\ Giuria 1, 10125 Torino, Italy
\and
INAF -- Istituto Nazionale di Astrofisica, Osservatorio Astrofisico di Torino, Strada Osservatorio 20, 10025 Pino Torinese, Italy
\and
Institut f\"ur Theoretische Physik, Universi\"at Heidelberg, Philosophenweg 16, 69120 Heidelberg, Germany
\and
Sub-department of Astrophysics, University of Oxford, Denys Wilkinson Building, Keble Road, Oxford, OX1 3RH, UK
\and
Division of Physics, Mathematics and Astronomy, California Institute of Technology, 1200 E California Blvd, Pasadena, CA 91125
}
\date{Received XXX, 2024; accepted XXX,2024}

\abstract
{}
{We cross-correlated galaxies from the LOw-Frequency ARray (LOFAR) Two-metre Sky Survey (LoTSS) second data release (DR2) radio source with the extended Baryon Oscillation Spectroscopic Survey (eBOSS) luminous red galaxy (LRG) sample to extract the baryon acoustic oscillation (BAO) signal and constrain the linear clustering bias of radio sources in LoTSS DR2.}
{In the LoTSS DR2 catalogue, employing a flux density limit of $1.5$ mJy at the central LoTSS frequency of 144 MHz and a signal-to-noise ratio (S/N) of $7.5$, additionally considering eBOSS LRGs with redshifts between 0.6 and 1, we measured both the angular LoTSS-eBOSS cross-power spectrum and the angular eBOSS auto-power spectrum. These measurements were performed across various eBOSS redshift tomographic bins with a width of $\Delta z=0.06$. By marginalising over the broadband shape of the angular power spectra, we searched for a BAO signal in cross-correlation with radio galaxies, and determine the linear clustering bias of LoTSS radio sources for a constant-bias and an evolving-bias model.}
{Using the cross-correlation, we measured the isotropic BAO dilation parameter as $\alpha=1.01\pm 0.11$ at $z_{\rm eff}=0.63$. By combining four redshift slices at 
$z_{\rm eff}=0.63, 0.69, 0.75$, and $0.81$, we determined a more constrained value of $\alpha = 0.968^{+0.060}_{-0.095}$. For the entire redshift range of $z_{\rm eff}=0.715$, we measured $b_C = 2.64 \pm 0.20$ for the constant-bias model,  $b(z)=b_C$, and then $b_D = 1.80 \pm 0.13$ for the evolving-bias model, $b(z) = b_D / D(z)$, with $D(z)$ denoting the growth rate of linear structures. Additionally, we measured the clustering bias for individual redshift bins.
}
{We detected the cross-correlation of LoTSS radio sources and eBOSS LRGs at a
9.2$\,\sigma$ statistical significance for one single redshift bin and 
at a 14.7$\,\sigma$ significance when the four redshift bins were combined. For 
the BAO signal, 
we achieved a significance of 2.2$\,\sigma$ for a single redshift bin, 2.7$\,\sigma$ for the combined cross-correlation and eBOSS auto-correlation, and 4$\,\sigma$ for the combined analysis of four redshift bins in the cross-correlation, when assuming a Gaussian distribution for the BAO dilation parameter.} 


\maketitle

\section{Introduction} \label{sec:Intro}

Since the discovery of the accelerated expansion of the Universe in 1998,  distinguished with the Nobel Prize in Physics in 2011 \citep{perlmutter1999measurements,riess1998observational}, there have been two popular explanations for this phenomenon. One relies on the presence of dark energy, a form of energy exhibiting negative pressure that acts in its simplest form as a cosmological constant ($\Lambda$) in the Einstein field equations (see e.g. \citealt{frieman2008dark} as a review). The other is based on modified gravity. The former relates to the substrate of the Universe, while the latter pertains to the dynamics of matter density perturbations. To test dark energy models, the observation of baryon acoustic oscillations (BAOs) provide one of the most robust probes of the Universe~\citep{percival2017baryon}. In the early Universe, acoustic waves spread in the primordial plasma with a characteristic speed of sound. At a redshift of $z \sim 1100$, photons decoupled from the baryonic matter as the Universe cooled, imprinting the characteristic scale of the acoustic horizon in the fluctuations of the cosmic microwave background \citep{dodelson:2003}. Due to the large entropy of the Universe, the decoupling of baryonic matter is slightly delayed and this falls into the potential wells of cold dark matter at the drag epoch of $z_\mathrm{d} \sim 1060$, at which the typical length of baryon acoustic density fluctuation $r_\mathrm{d} \sim 100 h^{-1}$ Mpc is fixed \citep{eisenstein1998baryonic}. This is why BAOs can serve as a `standard ruler' in the Universe. Overall, BAOs are immune to systematic uncertainties caused by the astrophysical properties of tracers and are only related to the kinematics of length scales, whereas the physics of BAOs is well understood. The quantities derived from the BAO length, the Hubble parameter, $H$, and the angular diameter distance, $D_A$, of galaxies at a given redshift can help us understand the nature of dark energy. From galaxy redshift surveys, we can derive both $H$ and $D_A$; without redshift information, we can derive $D_A$ as a function of redshift.

It is possible to detect BAOs  using a range of methods, each leveraging different tracers of the large-scale structure to probe the Universe’s expansion history. Galaxy clustering is the most widely used approach, whereby BAOs are manifested as a peak in the correlation function or oscillatory features in the power spectrum of galaxy distributions. This method was first used to detect BAOs in 2005 by the Sloan Digital Sky Survey (SDSS; \citealp{york2000sloan, eisenstein2005detection}) and the 2-degree Field Galaxy Redshift Survey (2dFGRS; \citealp{percival2007measuring}). Subsequent surveys, including  Baryon Oscillation Spectroscopic Survey (BOSS; \citealp{dawson2012baryon}),  extended BOSS (eBOSS; \citealp{dawson2016sdss}),  Dark Energy Spectroscopic Instrument (DESI; \citealp{aghamousa2016desi, desibao}), and  Dark Energy Survey (DES; \citealp{des1,des2}), have significantly improved the precision of BAO measurements. Cosmic microwave background (CMB) anisotropies provide another avenue for BAO detection, with the angular power spectrum of temperature and polarisation fluctuations encoding  early-Universe acoustic oscillations, as measured by Planck \citep{planck18} and Wilkinson Microwave Anisotropy Probe (WMAP, \citealp{hinshaw2013nine}). The Lyman-$\alpha$ forest, tracing neutral hydrogen through quasar absorption spectra, enables BAO detections at higher redshifts, as demonstrated by BOSS \citep{busca2012baryon, font2014quasar} and eBOSS \citep{des2020completed}. Intensity mapping, using the $ \text{H}\,\text{I} $ 21-cm line in particular, offers the potential to carry out BAO detection across vast cosmic volumes \citep{angulo2008detectability}. Multi-tracer methods, such as cross-correlating galaxies with the CMB lensing field or quasars with the Lyman-$\alpha$ forest, enhance BAO detection by mitigating cosmic variance \citep{font2014quasar, heymans2021kids}. Together, these methods highlight the versatility of BAOs as a cosmological probe, enabling precise measurements of the Universe’s expansion history and fundamental cosmological parameters.

Specifically, eBOSS is an optical spectroscopic survey conducted in the northern hemisphere as part of SDSS-IV. Building upon the success of its predecessor, BOSS \citep{dawson2012baryon}, eBOSS maps the  large-scale structure of the Universe by observing galaxies and quasars to improve our understanding of the Universe’s expansion history and cosmic distance scale. The catalogues offer data samples where observational systematics have been corrected and a selection function has been derived through random position sampling. The sources are categorised into three types: luminous red galaxies (LRGs), emission line galaxies (ELGs), and quasars (QSOs). In particular, the eBOSS-DR16  sample contains 174\,816 LRGs over a survey area of 4\,242 square degrees in the redshift interval of $0.6 < z < 1.0$ \citep{ross2020completed}. 

Meanwhile, with the forthcoming age of radio surveys, it will be feasible to probe the Universe to inherently higher redshifts than wide-area optical surveys can typically reach. The galaxies detected by optical surveys and radio surveys can then serve as tracers of the underlying large-scale structure. By cross-correlating optical and radio catalogues, we can extract information from optical surveys that radio surveys do not provide, such as radio source bias evolution and redshift distribution, as performed by  tools such as \texttt{Tomographer} \citep{chiang2019extragalactic}.

 LOw Frequency ARray (LOFAR; \citealp{van2013lofar}) is a radio interferometer array based in Europe, with a maximum baseline of approximately 2000 km and a resolution around or below 6 arcseconds. The array features two antennas: low band antenna (LBA) and high band antenna (HBA). The LOFAR Two-metre Sky Survey Data Release 2 (LoTSS DR2; \citealp{shimwell2022lofar}), which operates at a central frequency of 144 MHz using the HBA, provides the most extensive low-frequency radio catalogue to date, making it especially suitable for cosmological studies. With 841 individual pointings, LoTSS DR2 covers $\sim$5600 $\text{deg}^2$ over two regions and provides a total of $\sim$4.4 million radio sources.

While radio continuum surveys excel at probing high redshifts, they do not provide redshift information for individual sources. Consequently, cross-matching radio data with multi-wavelength surveys is essential for obtaining redshifts and host galaxy properties. Moreover, combining radio data with complementary datasets in different wavelengths (e.g. optical surveys or the CMB) not only constrains the overall redshift distribution of radio sources (e.g. \citealp{alonso2021cross}), but also helps verify that different tracers are mapping the same underlying dark matter structure. Due to the substantial 2D overlap between LoTSS DR2 and eBOSS, a corresponding 3D overlap is likely, which enables BAO extraction and cosmological parameter inference via cross-correlation. In addition, we can measure the bias of radio sources by cross-correlating them with optical sources. This cross-correlation will select a typical type of sources that share similar bias properties. For example, in this paper, we focus on the bias of radio sources that cross-correlate with optical luminous red galaxies. This paper is one in a series of cosmological analysis using LoTSS DR2, including the already published angular correlation function analysis (\citealp{h23}; hereafter H24) and cross-correlation with CMB lensing (\citealp{n23}; hereafter N24). The details of the radio source catalogue are stated in both H24 and N24.

This work is structured as follows. In Sect.~\ref{sec:data and methods}, we describe the catalogue we used for cross-correlation, the ways to construct the overdensity map and the survey mask, along with the formulae for theoretical and data auto- and cross-angular power spectrum $C_\ell$ values. 
In Sect.~\ref{sec:mock}, we present how to generate the mock catalogues, which are later used for pipeline testing, model comparison, and obtaining the covariance matrix of the measured 
$C_\ell$ values. The results are presented in Sect. \ref{sec:results}, including the BAO shift parameter and radio sources bias measurements. Finally, we discuss and present the conclusions of our work in 
Sect.~\ref{sec:conclusion}. Further details of the model selection 
are presented in Appendix~\ref{sec:apppen}. To summarise the key steps in our analysis, we provide a flowchart in Fig. \ref{fig:flowchart}, where the schematic illustrates the full pipeline from the data processing and power spectrum measurement to theoretical modelling, covariance estimation, and parameter inference. In this work, we chose the Planck 2018 best-fit cosmological parameters \citep{planck18} as the fiducial cosmology, namely $\Omega_\mathrm{c} h^2=0.120$, $\Omega_\mathrm{b} h^2=0.0224$, $n_\mathrm{s}=0.965$, $H_0=67.4 \mathrm{km}/\mathrm{s}/\mathrm{Mpc}$, and $\sigma_8=0.811$. 

It is worth noting that we adopted eBOSS LRG in the North Galactic Cap (NGC) to ensure a more homogeneous dataset. We do not incorporate other galaxy types observed by eBOSS, such as ELGs and QSOs. Due to the limited overlap of ELGs with LoTSS DR2 and the sparse distribution of QSO samples, these galaxy types are less valuable for cross-correlation. Additionally, different galaxy types exhibit distinct biases and redshift distributions \citep{bias-review}. In our theoretical calculations, we implicitly assume that the galaxy bias, $b(z)$, and the redshift distribution, $p(z)$, are identical across the survey footprint. Therefore, we select only one type of galaxy, the LRGs, which contains the most cross-correlation information. Throughout this paper, when we refer to `redshift binning' in the angular power spectrum, we specifically mean the redshift bins used in eBOSS.

\begin{figure*}
\centering
\includegraphics[width=1\hsize]{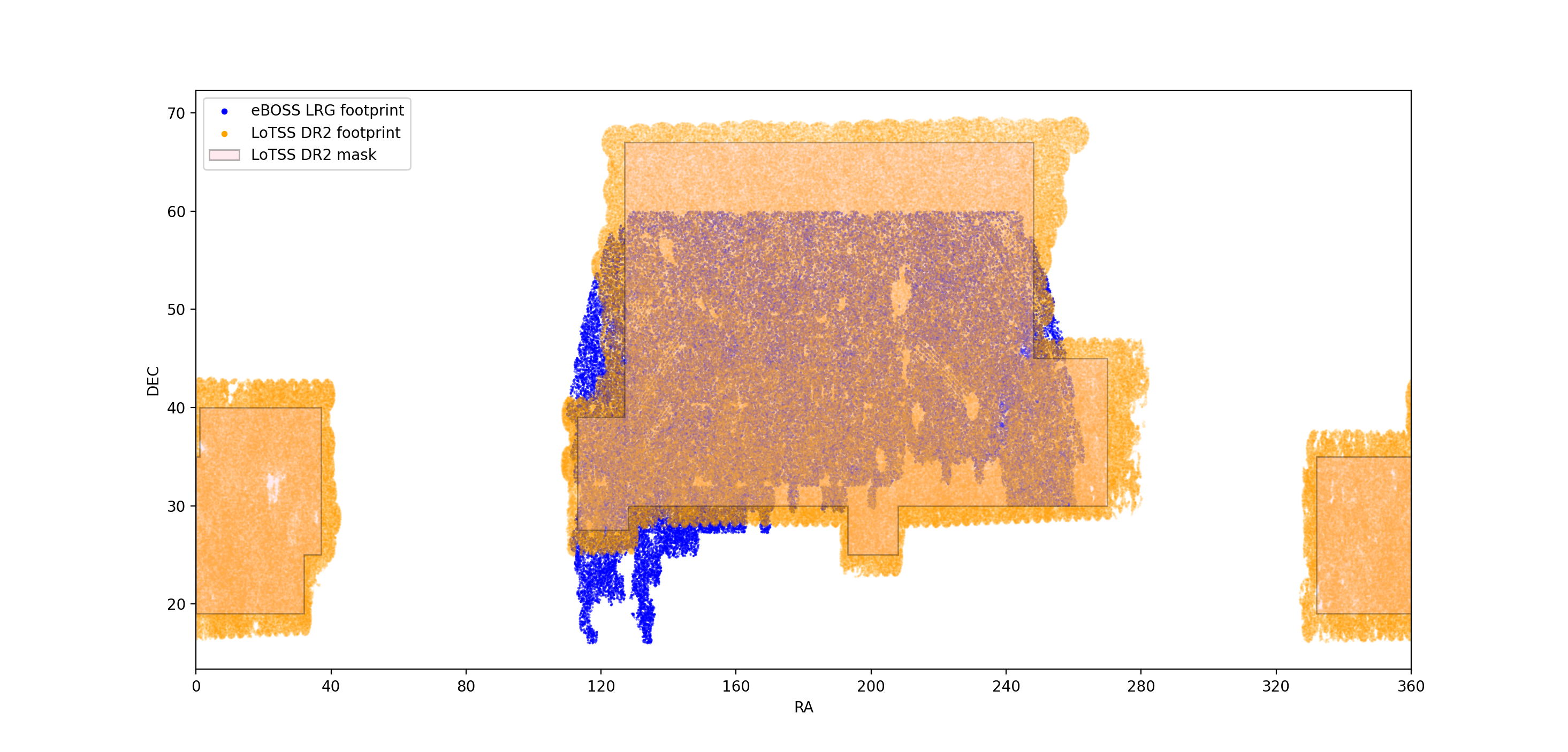}
   \caption{Footprint covered by the LoTSS DR2 (in orange) and eBOSS LRGs NGC(in blue), plotted against right ascension (RA) and declination (DEC). This visualisation highlights the significant overlap between the two surveys, which allows for a cross-correlation analysis. The mask  applied on the LoTSS DR2 data (in pink shaded region with black edge) is also shown.}.
      \label{fig:footprint}
\end{figure*}

\begin{figure*}
    \centering
    \includegraphics[width=0.8\hsize]{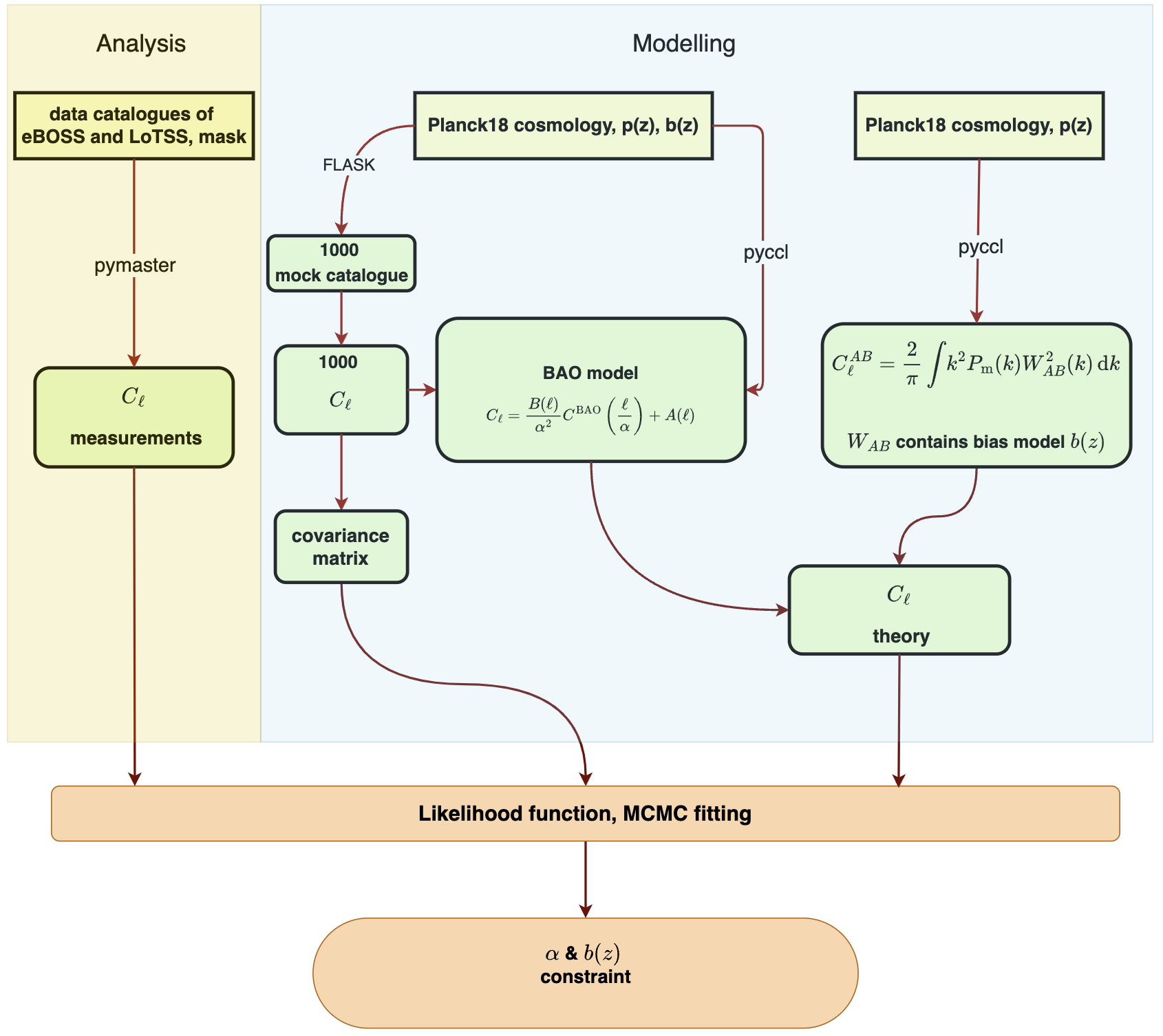}
    \caption{Flowchart of this work. Left panel: Analysis (in light yellow region), representing the process of using real survey data to measure the angular power spectrum. Right panel: Modelling (in light blue region), which consists of generating mock catalogues and computing the theoretical angular power spectrum. The measured $C_\ell$, the theoretical prediction, and the covariance matrix (computed from 1000 mock catalogues) are then used to construct the final likelihood. The likelihood is subsequently used for Markov chain Monte Carlo (MCMC) parameter estimation. Note: the theoretical templates differ depending on whether the constraints are on the BAO parameters or the bias parameters.}
    \label{fig:flowchart}
\end{figure*}

\section{Data and methodology} \label{sec:data and methods}

\subsection{Data: LoTSS DR2 radio sources and eBOSS LRGs}

LoTSS DR2 covers approximately 28\% of the northern hemisphere, while eBOSS LRGs cover about 15\%. As described in Sect.~\ref{sec:Intro}, LoTSS is a radio continuum survey with the central frequency around 144 MHz. Since radio continuum measurements average over all spectral information, the continuum data do not provide redshift information for radio sources. However, the angular position information contained in the LoTSS DR2 catalogue, specifically the right ascension (RA) and declination (DEC), is suitable for cosmological analysis. On the other hand, the eBOSS LRGs catalogue contains both angular coordinates and spectroscopic redshifts. As illustrated in Fig. \ref{fig:footprint}, approximately 50\% of the area in LoTSS DR2 overlaps with eBOSS LRGs, and 92\% of the eBOSS LRG area is covered by LoTSS DR2.  The redshift range of eBOSS LRGs is \(0.6 < z < 1\) (shown in Fig.~\ref{fig:nzbz}), as it is a spectroscopic survey. For the redshift distribution of LoTSS DR2 sources, we used LOFAR Deep Fields to represent LoTSS (see \citealt{ken21} and its use in H24). Given the large angular overlap (as shown in Fig. 1) and the partial radial overlap indicated by the redshift distribution overlap in Fig. \ref{fig:nzbz}, it is natural to expect a volume overlap in the three-dimensional space. Therefore, performing cross-correlation should be viable. Cross-correlations can reduce certain systematic effects compared to auto-correlation measurements, as systematics between two independent surveys are unlikely to correlate. However, shot noise does contribute to the cross-power spectrum due to the finite number of cross-pairs. This contribution is naturally included in our error budget (derived from mock catalogues). These mocks capture observational conditions and statistical uncertainties and provide a robust estimate of the overall errors in our measurements. While shot noise does increase the variance of the cross $C_\ell$ measurements, it does not affect the amplitude, making cross-correlation a powerful tool for obtaining cleaner and more accurate measurements. 

Our mask imposed on the LoTSS DR2 radio source catalogue\footnote{LoTSS DR2 catalogue: \url{https://lofar-surveys.org/dr2_release.html}} is a weighted mask where each pixel is assigned a weight to account for systematics over the footprint. We summarise the construction of the mask as follows: Firstly, random positions were generated within the survey area, correcting for smearing effects. Next, we adjusted detection likelihood using completeness as a function of signal-to-noise ratio (S/N) and flux density and derived `measured' flux densities for random sources. Positional cuts were applied to handle systematics, and inner regions were masked accordingly. Finally, random sources were validated against data distributions, and we applied an S/N cut of 7.5 and a flux density cut of 1.5 mJy. Details are provided in H24 and N24.

For the eBOSS LRG catalogue\footnote{eBOSS LRG catalogue: \url{https://data.sdss.org/sas/dr16/eboss/lss/catalogs/DR16/}}, we used the RA, DEC, redshift, and weight from both the survey catalogue and the random catalogue to construct the overdensity map and survey mask. Their details are described in the following section.

\subsection{Methods}
\subsubsection{The measured $C_\ell$ }

To quantify the galaxy number density fluctuations at different scales and to link them to cosmological models, a common method is to measure the power spectrum in harmonic space or the correlation function in configuration space. In this work we use the former (dubbed $C_\ell$), where $\ell$ is the angular multipole. To measure the $C_\ell$ of the cross correlation between eBOSS LRGs and LoTSS DR2, as well as their respective auto-correlation power spectra, we use the python package \texttt{pymaster}~\citep{alonso2019unified}\footnote{\url{https://namaster.readthedocs.io/en/latest/}}. It was originally designed to analyse the angular power spectrum of the CMB temperature and polarisation in the \texttt{Healpix} \citep{gorski1999healpix} tessellation; it later included the effects of the limited survey geometry that causes the coupling between different multipoles. In \texttt{pymaster}, we need to input the overdensity map and the mask of the survey and output the $C_\ell$ values. We note that the map and mask in eBOSS and LoTSS are constructed slightly differently.

Regarding LoTSS, the catalogues include the angular positions of the radio sources. Sources in each pixel are assigned a weight, as described in the mask description. This weight quantifies the fraction of the sources that should have been observed compared to the real underlying number of sources, accounting for survey systematics and instrumental effects.

On the other hand, in the eBOSS LRG catalogue, each individual source is assigned a weight: \( w = w_\mathrm{FKP}w_\mathrm{systot}w_\mathrm{noz}w_\mathrm{cp} \). The weight, \( w_\mathrm{FKP} \), is radially dependent and minimises the variance of the 3D power spectrum measurement. Following \citealp{alonso2021cross}, we neglected this weight because we are computing the angular power spectrum, not a 3D power spectrum. The other weights depend on direction: \( w_\mathrm{systot} \) corrects for spurious fluctuations in the photometric target selection, \( w_\mathrm{noz} \) accounts for redshift failures, and \( w_\mathrm{cp} \) addresses fiber collisions. These weights are provided by the publicly available eBOSS catalogues. The eBOSS weights per pixel are the average weights of the sources per pixel and are, thus, are essentially the ratio of the underlying true number of sources per pixel with respect to the observed ones. The mask of eBOSS is also weighted, constructed using the averaged weights per pixel from its random catalogue.

Hereafter we refer $\mathrm{L}$ to 'low nside' $n_\mathrm{side}=512$ and $\mathrm{H}$ to 'high nside'  $n_\mathrm{side}=4096$ in the \texttt{Healpix} regime, survey A, as the LoTSS DR2 survey and survey B is the eBOSS survey. As described above, the overdensity map, $\delta_p$, for LoTSS in both auto-correlation and cross-correlation is expressed as:\ 
\begin{equation}
\label{eq:delta_A}
    \delta_p^A = \frac{n_p^{\mathrm{L}}}{\bar{n}^{\mathrm{L}}}-1,
\end{equation}
where $n_p$ is the weighted number of radio sources in each pixel and $\bar{n}$ represents the weighted mean number of radio sources per pixel. For a weighted mask $w_p$,  $n_p = N_p/w_p$ and $\bar{n}=\sum_p n_p/\sum_p w_p$, 
following the method of optimal inverse-variance weighting, where $N_p$ is the source count in each pixel.\footnote{
  Our estimator for the weighted mean,
  \begin{equation}\label{eq:optimal-weight}
    \bar{n} = \frac{\sum_{p} N_{p}}{\sum_{p} w_{p}},
  \end{equation}
  is equivalent to the standard inverse-variance weighted mean under the assumption of Poisson statistics. 
  Specifically, if $N_{p}$ follows a Poisson distribution, then $\mathbb{E}[N_{p}] = \mathrm{Var}(N_{p})$, 
  and defining $n_{p} = N_{p}/w_{p}$ implies 
  \begin{equation}
    \mathrm{Var}\bigl(n_{p}\bigr)
    = \mathrm{Var}\Bigl(\frac{N_{p}}{w_{p}}\Bigr)
    = \frac{1}{w_{p}^{2}}\, \mathrm{Var}(N_{p})
    = \frac{1}{w_{p}}\, \mathbb{E}\bigl(n_{p}\bigr)
    \;\propto\;\frac{1}{w_{p}}.
  \end{equation}
  Thus, the appropriate inverse-variance weight is $w_{p}$. Substituting 
  $x_{p} = n_{p}$ and $\sigma_{p}^{2} = 1/w_{p}$ into the usual formula,
  \begin{equation}
    \bar{x} = \frac{\sum_{p} \bigl(x_{p}/\sigma_{p}^{2}\bigr)}{\sum_{p} \bigl(1/\sigma_{p}^{2}\bigr)},
  \end{equation}
  we recovered Eq. \ref{eq:optimal-weight},
  which matches the optimal inverse-variance weighted average. 
}

The overdensity map $\delta_p^B$ for eBOSS in cross-correlation is defined the same way as in Eq.~(\ref{eq:delta_A}); whereas in the eBOSS auto correlation, it is expressed as:\ 
\begin{equation}
    \delta_p^B = \frac{n_{p,d}^{\mathrm{H}} - \beta n_{p,r}^{\mathrm{H}}}{\beta  n_{p,r}^{\mathrm{L2H}}}.
\end{equation}
$\mathrm{L2H}$ in the denominator denotes an upgrade from low nside to high nside, $d$ is data catalogue, and $r$ the random catalogue. Then, $\beta = \sum_p n_{p,d}/n_{p,r}$ quantifies the ratio of the total number of random sources to the total number of galaxy sources. Upgrading from low-resolution to high-resolution map allows us to compute the shot noise by fitting the flat tail of the eBOSS auto-correlation $C_\ell$ at high $\ell$, as there is no explicit formula for the shot noise term when each individual galaxy is weighted. For more details of this upgrading method, we refer to \cite{garcia2021growth}.

With the above maps and masks, we calculate the $C_\ell$ measurements using the pseudo-$C_\ell$ estimator \citep{1973ApJ...185..413P} applied in \texttt{NaMaster}\citep{alonso2019unified}, considering the mode coupling due to partial sky surveys, the shot noise term for auto angular power spectrum due to discrete galaxy number density, and band powers. 

We computed the covariance of the measured $C_\ell$ using $1000$ mocks generated by {\tt FLASK} \citep{xavier2016improving}, assuming lognormal fields on the sphere.  The mock catalogues also serve as a pipeline test for our $C_\ell$ measurements. Details of the mocks are described in Sect. \ref{sec:mock}. Given that about 5\% of galaxies in the LoTSS DR1 value-added catalogue \citep{wendy-dr1-value} exhibit multi-component effects that might affect the amplitude of the angular power spectrum \citep{blake2004angular}, we assigned up to 25\% of the galaxies in the radio mock catalogue as double or triple sources at the same position. We found no evidence that the differences in cross $C_\ell$ measurements with and without multi-component injection are significant. This is believed to be due to the lack of correlation of systematics in cross-correlation as well as the narrow redshift bin of eBOSS being applied, which largely reduces the number of potential multi-component galaxies in the radio catalogue that could be cross-correlated with optical galaxies.

\begin{figure}
    \centering
    \includegraphics[width=\hsize]{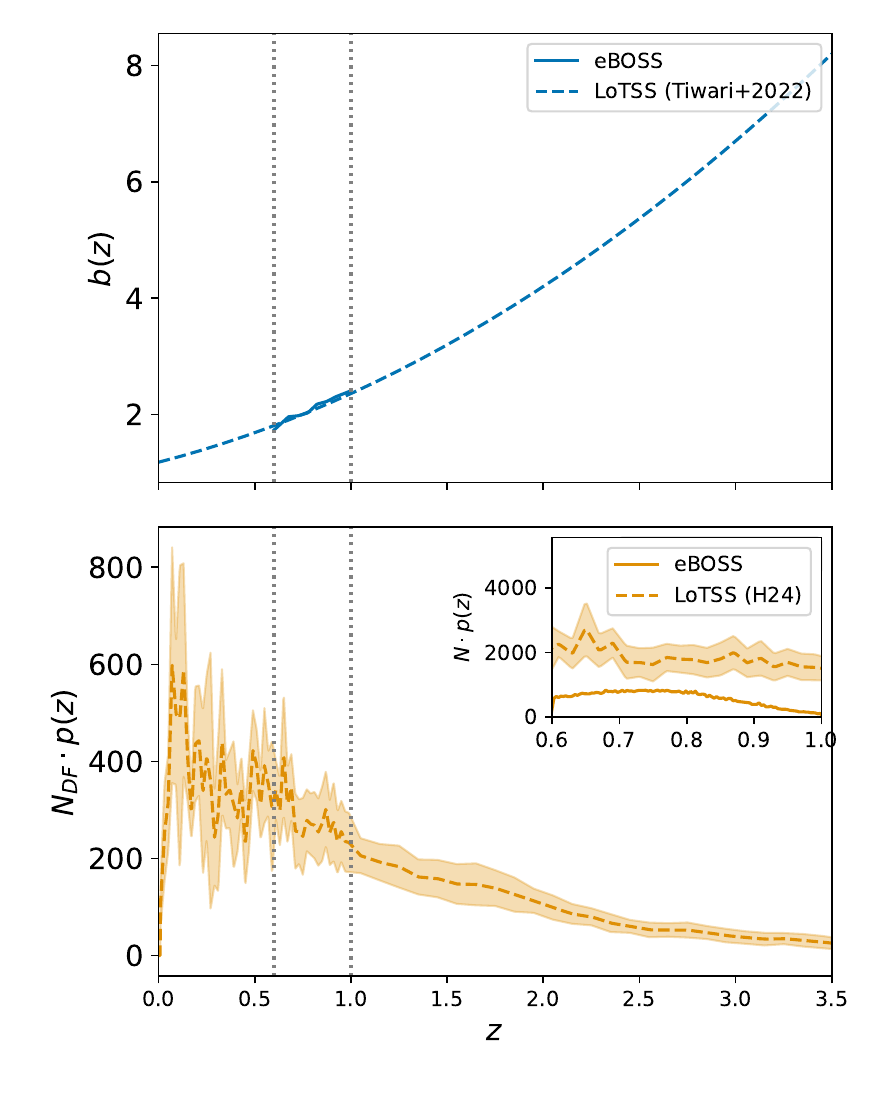}
    \caption{Upper panel: Model prediction of the bias function $b(z)$ for eBOSS and LoTSS DR2 surveys. The vertical dotted lines show the redshift range of eBOSS LRGs. Lower panel: Redshift distribution of LoTSS radio sources as inferred from the exquisite multi-wavelength coverage of LoTSS Deep Fields (see \citealt{ken21, h23, bhardwaj2024flux} for details). $N_{DF}$ represents the total number of galaxies in LOFAR Deep Fields and $p(z)$ is the normalised redshift distribution. The inset shows the redshift distribution $N p(z)$ for eBOSS and wide field LoTSS DR2 surveys, where the $N$ denotes the total number of galaxies in each survey, respectively. The histogram redshift interval is 0.0035. The $b(z)$ and $p(z)$ shown here are used as inputs to generate the mock catalogue.}
    \label{fig:nzbz}
\end{figure}

\subsubsection{The theoretical $C_\ell$}
\label{sec:theory}

We computed the theoretical angular power spectrum using \texttt{pyccl} \citep{Chisari_2019}\footnote{\url{https://github.com/LSSTDESC/CCL}}. We summarise the calculation process as follows.
First, the observed extragalactic sources are projected on the sphere of the sky, and the theoretical 2D overdensity map $\Delta_\mathrm{g}(\mathbf{\hat{r}})$ can be linked to the 3D density contrast of matter, $\delta_\mathrm{m}(\bf{r})$, and the galaxy clustering bias, $b(z)$. We start from the definition of 3D galaxy density contrast and its relation to matter density contrast:
\begin{eqnarray}
\delta_\mathrm{g}(\mathbf{r})= b(z) \delta_\mathrm{m}(\mathbf{r}),
\end{eqnarray}
The 2D density contrast can be expressed as an integral of the 3D density contrast along the line of sight:
\begin{eqnarray}
\Delta(\mathbf{\hat{r}})= \frac{n(\mathbf{\hat{r}})}{\bar{n}} - 1 = \int\! \delta_\mathrm{g}(\mathbf{r}) p(r)\, \mathrm{d} r, 
\end{eqnarray}
where $n(\mathbf{\hat{r}})$ is the projected source number density per steradian in direction, $\mathbf{\hat{r}}$, and $\bar{n}$ represents the mean source number density per steradian; $p(r)$ is the distribution of sources as a function of distance, $r$, convoluted with the sensitivity of the survey, which in turn is a 
function of cosmological redshift, $r = r(z)$.  

The theoretical galaxy angular power spectrum of two surveys A and B is expressed as:\ 
\begin{equation}
\label{eq:pk2cl}
C_\ell^{A B}=\frac{2}{\pi} \int\! k^{2} P_\mathrm{m}(k) W_{AB}^{2}(k)\, \mathrm{d} k ,
\end{equation}
where $P_\mathrm{m}(k) = |\delta_\mathrm{m}(k)|^2$ is the matter power spectrum and the window function of two surveys \(W_{A B}(k)\) is 
\begin{equation}
\label{eq:WAB}
W_{A B}(k) \!=\!\!\! \int\!\!\! D(z) b_{A}(z) p_{A}(r) j_\ell (k r)\, \mathrm{d} r 
\!\! \int\!\!\! D(z) b_{B}(z) p_{B}(r) j_\ell(k r)\, \mathrm{d} r.
\end{equation}

We chose the a redshift interval of 0.06 at $z\sim 0.6$ and $\ell>50$ for angular scale, thus the window function (or the 'kernel') is sufficiently broad compared to the angular correlation length. This enables us to apply the Limber approximation \citep{limber1}. The spherical Bessel function $j_\ell(x)$ is approximated by a Dirac delta function $\delta$, centered at the first peak around $kr = \ell + \frac{1}{2}$, leading to the substitution $j_\ell(x) \mapsto \sqrt{\frac{\pi}{2 \ell + 1}} \delta \left( \ell + \frac{1}{2} - x \right)$, thereby simplifying and accelerating the computation of the theoretical $C_\ell$s \citep{limber2}.
 The redshift interval and the choice of $\ell$ limits have been validated by our mock catalogue comparing with what is obtained in \texttt{CAMB} \citep{lewis2011camb}  without assuming a Limber approximation.

In this paper, for the BAO signal fitting (i.e.  aimed at measuring the angular diameter distance $D_A(z)$ via the shifting parameter $\alpha$), we used a template that mimics the nonlinear evolution of the BAO, while marginalising over the broadband shape of the angular power spectrum that is caused by galaxy bias, nonlinear matter evolution, large- and small-scale redshift space distortions (RSDs), and other cosmological parameters.

The model for the eBOSS auto angular power spectrum \citep{seo2012bao,des1,des2,anderson2014clustering,gil2016clustering,ross2017optimized,ata2018clustering,zhao2019clustering,bautista2018sdss} is:\ 
\begin{equation}\label{bao_model}
    C_\ell =  \frac{B(\ell)}{\alpha^2}C^\mathrm{BAO}\left(\frac{\ell}{\alpha}\right)+A(\ell). 
\end{equation}
We adopted the same structure for the model of the cross angular power spectrum, but allowing the expressions $B(\ell)$ and $A(\ell)$ to differ.
The BAO dilation parameter $\alpha$ measures the angular location of the BAO observed (obs) relative to that of the fiducial cosmology (fid):
\begin{equation}
\alpha (z)=\left[D_A(z) / r_\mathrm{d}\right]_{\mathrm{obs}} /\left[D_A(z) / r_\mathrm{d}\right]_{\mathrm{fid}}, 
\label{eq: alpha-def}
\end{equation}
where $D_A(z)$ represents the angular diameter distance, and $r_\mathrm{d}$ is the sound horizon scale at the drag epoch. Any deviation from $\alpha=1$ indicates a deviation of 
the true cosmology from the fiducial one. 

The factor $B(\ell)/\alpha^2$ in Eq.~(\ref{bao_model}) is due to the transformation of the solid angle 
$(\theta/\theta_\mathrm{fid})^2 = \alpha^{-2}$ (see also Eq. 16 in \citealt{camera2022novel}). Compared with the $B(\ell)$ alone, it is shown to reduce the errors on the constraints of the nuisance parameters $B(\ell)$ and the degeneracy between $\alpha$ and $B(\ell)$ . The template $C^\mathrm{BAO}$ is a projection of $P^\mathrm{BAO}(k)$ using 
Eq.~(\ref{eq:pk2cl}); the 3D BAO template $P^\mathrm{BAO}(k)$ is expressed as:\ 
\begin{equation} \label{eq:nw}
P^\mathrm{BAO}(k) = \left[P^{\mathrm{lin}}(k) - P^{\mathrm{nw}}(k)\right] \exp\left(-\frac{k^2 \Sigma_{nl}^2}{2}\right) + P^{\mathrm{nw}}(k),
\end{equation}
where $P^\mathrm{lin}(k)$ is the linear power spectrum, $P^\mathrm{nw}$ is the power spectrum without BAO (denoted as 'no-wiggle' as per \citealp{eihu1998}), and $\Sigma_{nl}$ is the nonlinear damping parameter. We fixed $\Sigma_{nl}=5.5\ \mathrm{Mpc}/h$ as used in eBOSS LRG \citep{bautista2018sdss}. We  tested the range of $2-8\ \mathrm{Mpc}/h$ using the mock and found that the impact on the cross-correlation is insignificant; in fact, the smoothing of the BAO is dominated by the width of the underlying redshift distribution, while the exact choice of the non-linear damping parameter does not have a prominent impact \citep{seo2012bao}.

\begin{figure}
\centering
\includegraphics[width=\hsize]{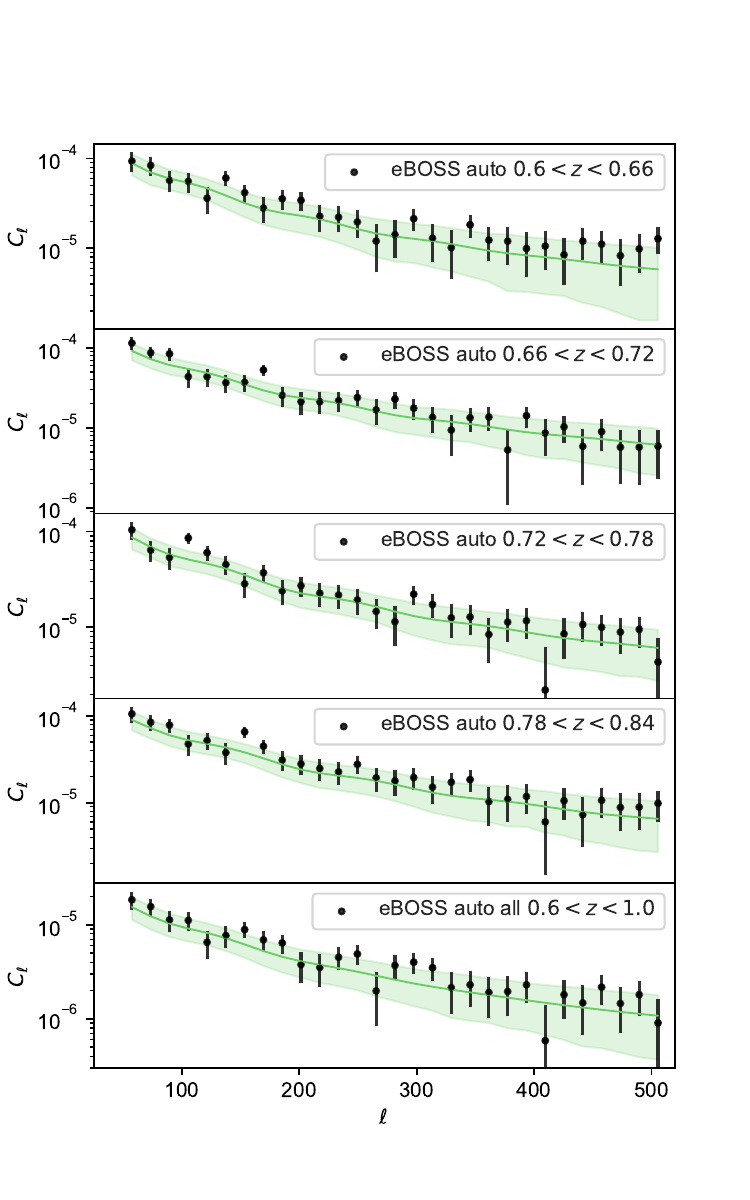}
   \caption{ eBOSS auto-angular power spectrum computed from mock and real catalogue in different redshift bins. Presented green curves are the mean $C_\ell$ values of the 1000 mocks, the black dots with errorbars represent the measured $C_\ell$ from the real catalogue. The errors are obtained from the square root of the diagonal elements of the covariance matrices calculated using the mock.}
      \label{fig:cl-auto-mock}
\end{figure}

\begin{figure}
\centering
\includegraphics[width=\hsize]{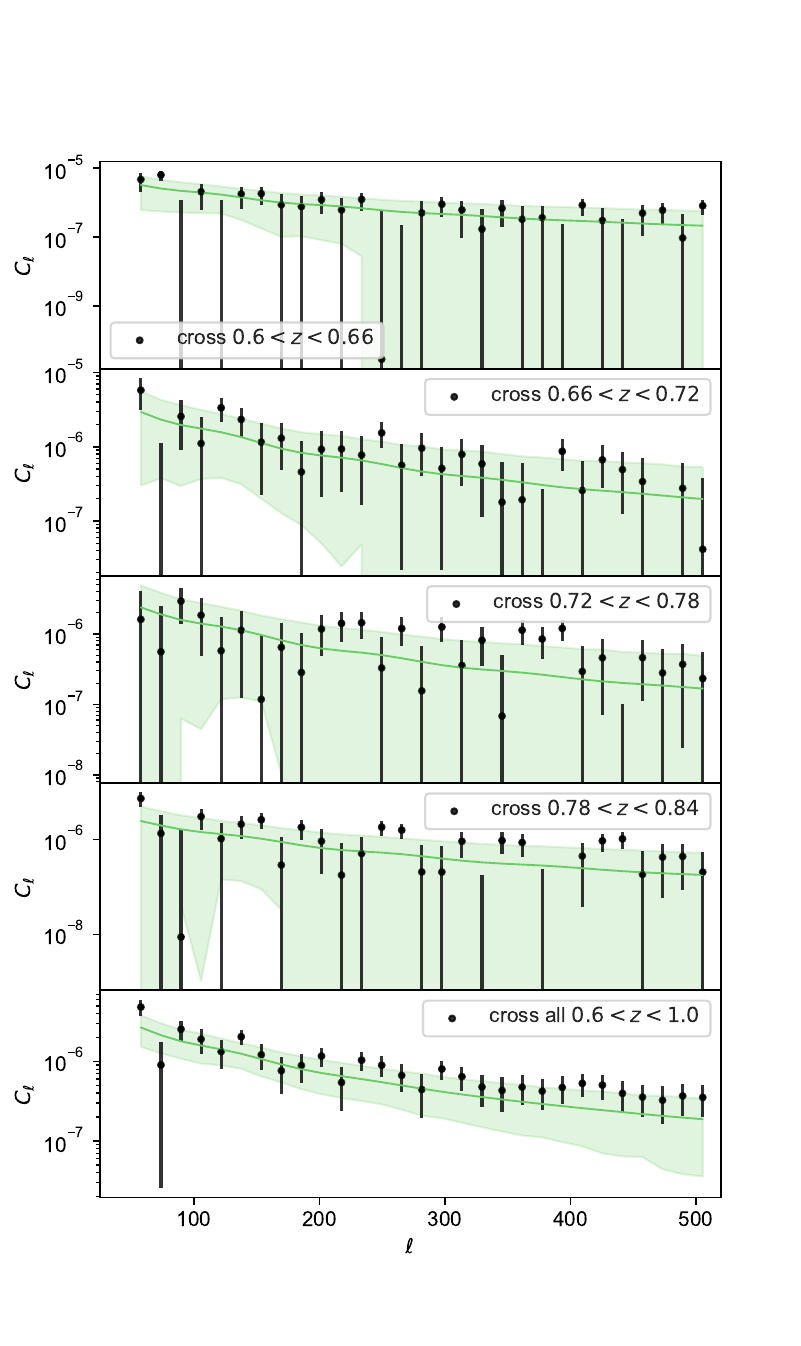}
   \caption{ LoTSS and eBOSS cross-correlation mean $C_\ell$ values (green curves) from the mocks are contrasted with the measured $C_\ell$ values from the real catalogue (black points with error bars) in different redshift bins. The error bars indicate measurement uncertainties, which are derived from the covariance matrices computed from 1000 mocks.}
      \label{fig:cl-cross-mock}
\end{figure}

\begin{figure*} 
   \centering
   \includegraphics[width=\textwidth]{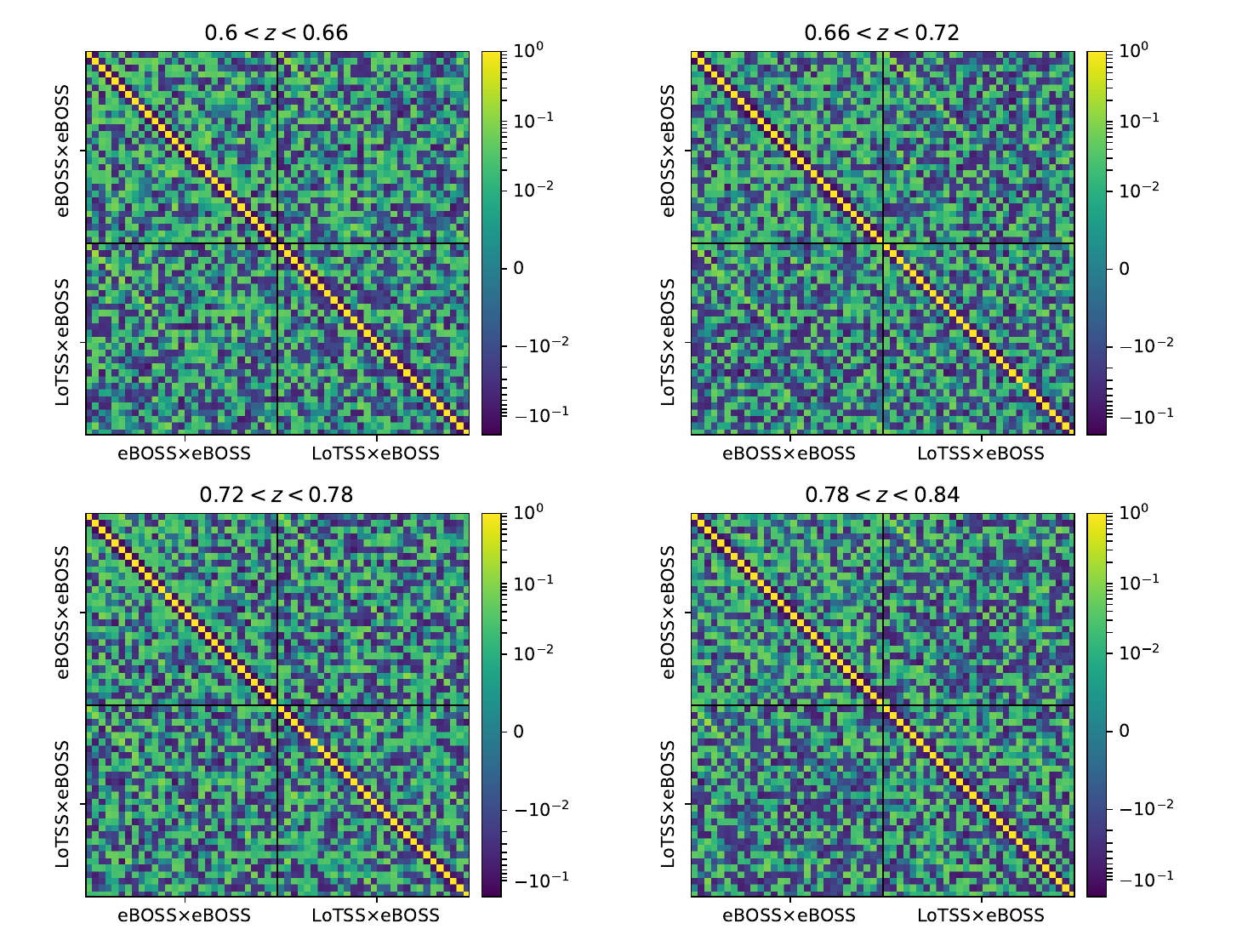}
    \caption{ Correlation matrices for the eBOSS auto-correlation and LoTSS$\times$eBOSS cross-correlation angular power spectrum measured from the mock catalogues in four redshift slices shown in the legend. In each block, the angular scale interval is $\Delta_\ell = 16$ in the range of $50<\ell<500$.}
         \label{fig:corr4zbin}
\end{figure*}

\begin{figure}\label{corrmat_all}
\centering
\includegraphics[width=0.8\hsize]{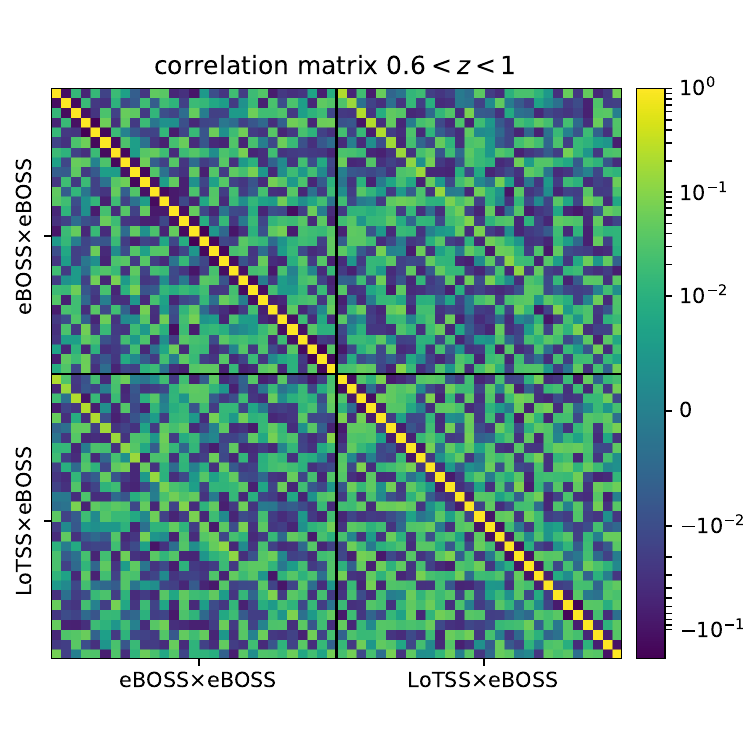}
   \caption{ Correlation matrix for the eBOSS auto-correlation and LoTSS$\times$eBOSS cross-correlation angular power spectrum measured from the mock catalogues in the redshift range of $0.6<z<1$. In each block, the angular scale interval is $\Delta_\ell = 16$ in the range of $50<\ell<500$.}
      \label{fig:corrallzbin}
\end{figure}

After the mock test conducted in Sect.~\ref{sec:mock}, we adopted the fitting range $50 < l < 500$ and the scale interval $\Delta \ell=16$, ensuring that our scale interval is smaller than one period of the BAO wiggles according to the Nyquist theorem. A polynomial function (w.r.t. which fitting function to be used, refer to \citealt{seo2008nonlinear}) was used for $B(\ell)$ and $A(\ell)$ for both the auto-correlation and cross-correlation, based on the mock tests in Sect.~\ref{sec:mock} and in 
Appendix~\ref{sec:apppen}:

\begin{equation}\label{eq:defaultBAOtemplate}
    B(\ell) = B_0; \quad A(\ell) = a_{1}+a_{2}/\ell+a_{3}\ell  .
\end{equation}

Thus, for the BAO analysis, we have five parameters to be fitted $(\alpha, B_0, a_1, a_2, a_3)$, where $ \alpha $ is the main parameter and the others are nuisance parameters.
For the measurement of LoTSS DR2 bias, we utilised Eq.~(\ref{eq:pk2cl}) and treated $b(z)$ in the window function (Eq.~\ref{eq:WAB}) as a free parameter to be estimated. We primarily consider two models for $b(z)$:

\begin{itemize}
    \item Constant bias model: This model assumes $b(z) = b_C$, where $b_C$ is a constant.
    \item Evolving bias model: Assuming $b(z) = b_D/D(z)$, where $b_D$ is a constant and $D(z)$ represents the linear growth factor at a given redshift.
\end{itemize}

In both bias models, there is one parameter, $b_C$ or $b_D$ to estimate.
We used the cross-correlation $C_\ell$ to measure the bias,  assuming that the statistical errors and systematics between different surveys are not correlated; there is no need to subtract the noise term, which is one of the advantages of using cross-correlation as a probe.

With the above measured and theoreical $C_\ell$ values, to estimate the cosmological parameters from the observed data, we employed a likelihood function based on the comparison between the theoretical and observed angular power spectra. The likelihood is constructed as a Gaussian function of the form:

\begin{equation}
\mathcal{L} \propto \exp\left(-\frac{1}{2} \Delta C_\ell^T \mathbf{C}^{-1} \Delta C_\ell\right),
\end{equation}

where \(\Delta C_\ell = C_\ell^{\text{data}} - C_\ell^{\text{theory}}\) represents the difference between the data and theory power spectra and \(\mathbf{C}^{-1}\) is the inverse of the covariance matrix of the data. The construction of the covariance matrix, which accounts for the statistical uncertainties and correlations between different angular scales, \(\ell\), is detailed in Sect. \ref{sec:mock}. By minimising this likelihood, we were able to obtain the best-fit parameters to describe our observations. Here, $\chi^2$ is defined as $-2\log\mathcal{L}$.

\subsection{Choice of theoretical $b(z)$ and $p(z)$} \label{sec:choose_bz_nz}
In calculating the theoretical angular power spectrum, we projected the 3D power spectrum into 2D along the line of sight, as indicated by Eqs.~(\ref{eq:pk2cl}) and (\ref{eq:WAB}). For this process, assumptions regarding the bias function $b(z)$ and the redshift distribution $p(z)$ are necessary for both surveys.

We list the options for $b(z)$ and $p(z)$ as below and show them in Fig.~\ref{fig:nzbz}:
   \begin{itemize}
      \item $b(z)$ for LoTSS DR2: Utilising the LoTSS DR1 best fit result from \cite{tiwari2022galaxy}, expressed as $b(z)=1.18 + 0.85z + 0.33z^2$.
      \item $p(z)$ for LoTSS DR2:  LOFAR Deep Fields \citep{h23}
      \item $b(z)$ for eBOSS LRGs: Derived from the effective Zel’dovich approximation mock for eBOSS (EZmock, \citealp{ezmock}, originally from \citealp{chuang2015ezmocks}).
      \item $p(z)$ for eBOSS LRGs: Directly obtained from galaxy redshift catalogue.
 
   \end{itemize}

Regarding the $b(z)$ for eBOSS LRGs, we calculate the power spectrum monopole, denoted as $P_\mathrm{g}(k) = b^2 P_\mathrm{m}(k)$, from 1000 EZ mocks using {\tt nbodykit} \citep{nbodykit}. The bias values ($b$) are obtained by taking the ratio of the galaxy power spectrum $P_\mathrm{g}(k)$ to the matter power spectrum $P_\mathrm{m}(k)$. We observe that the bias values obtained are approximately 1.3 times lower than the approximation from \citet{dawson2016sdss}, expressed as $1.7 \sigma_{8,0} / \sigma_8(z)$. This difference reflects the use of two complementary approaches: the \citet{dawson2016sdss} model provides a general framework for understanding clustering trends, while our mock-based approach incorporates survey-specific details and observational conditions in the final eBOSS sample. Additionally, a comparison of the eBOSS auto angular power spectrum of EZ mock test results and the correlated mock is crucial and the theoretical input must remain consistent.

The redshift distribution in LoTSS DR2 is significantly broader compared to eBOSS, covering approximately $0 < z < 7$ in its range\footnote{Note that in \citealp{ken21}, the redshift fitting only extended to $z = 7$; this is therefore a modelling limit, not necessarily a true limit.}. As shown in Fig.~\ref{fig:nzbz}, we restrict the redshift range for our LoTSS mock catalogue to $0 < z < 3.5$. This is because the eBOSS catalogue, which covers a narrow redshift slice of $0.6 < z < 1.0$, contains far fewer galaxies at higher redshifts, making cross-correlation unlikely. We note that in BAO analysis, the bias we assume in LoTSS originates from the measurements in LoTSS DR1 \citep{tiwari2022galaxy}. While this may not accurately represent the bias in LoTSS DR2, it will not affect the analysis because bias effects in the angular power spectrum are part of the broadband shape and will be marginalised out. Therefore, the measured $C_\ell$ using the LoTSS DR2 catalogue may slightly deviate from the input theoretical $C_\ell$ in our mock catalogue, as shown in Fig.~\ref{fig:cl-cross-mock}. In our radio source bias measurement, where bias is treated as a parameter to be fitted using the data, we take a theoretical bias of 1 and do not assume any LoTSS bias in the theoretical model.

\section{Mock test} \label{sec:mock}
 
 We carried out mock tests for three main purposes: pipeline validation, generating the covariance matrix of $C_\ell$s, and testing the robustness of the BAO template parametrisation. To achieve this, we employed the Full-sky Lognormal Astro-fields Simulation Kit (\texttt{FLASK}; \citealp{xavier2016improving}) to generate 1000 sets of mock catalogues for both eBOSS LRGs and LoTSS DR2. Specifically, this involves generating 1000 mocks for eBOSS and 1000 mocks for LoTSS DR2. Importantly, a pair of LoTSS DR2 and eBOSS mocks is generated using the same seed, ensuring correlation between them. To replicate the LoTSS DR2 and eBOSS LRG catalogs, we create log-normal density fields tomographically using FLASK. In particular, the LoTSS galaxies are generated in the redshift range $0$ to $3.5$, and eBOSS LRGs are generated in the redshift range $0.6$ to $1.0$.

All statistical properties, including auto and cross-correlations, are determined by the input angular power spectrum. Additionally, effects such as redshift-space distortions and lensing are included in the simulation via the input spectra provided to the FLASK pipeline. The input theoretical angular power spectrum, $C_\ell$, in different redshift bins, is generated using \texttt{CAMB}. This includes considerations for the observed number counts of RSD, non-linear power spectrum corrections, and lensing. The redshift distribution profile, $N(z) \equiv N p(z)$ is provided to \texttt{FLASK} along with $C_\ell$ to generate mock number count maps.

\subsection{Pipeline test and covariance of auto- and cross-angular power spectrum}

The mock catalogue is first used for pipeline validation that will be applied to the real data in Sect. \ref{sec:results}. Figure~\ref{fig:corr4zbin} shows the mock test results of the eBOSS auto angular power spectrum and cross power spectrum at four eBOSS redshift bins, $0.6<z<0.66$, $0.66<z<0.72$, $0.72<z<0.78$, and $0.78<z<0.84$, while Fig.~\ref{fig:corrallzbin} shows both of the angular power spectra combining all redshift bins $0.6<z<1.0$. We chose the redshift binning $\Delta_z=0.06$ because the redshift resolution in our mock is $\Delta_z=0.02$ and the Nyquist frequency is within the range of the wavelength of a BAO wiggle in the power spectrum \citep{condon2016essentialradio} . These are also shown in the errors of the measured $C_\ell$s in Figs.~\ref{fig:cl-auto-mock} and \ref{fig:cl-cross-mock}, which are the diagonal elements of our convariance matrix computed from 1000 mock $C_\ell$s. The way to calculate the elements of the covariance matrix between two multipoles $\ell_1$ and $\ell_2$ is the following:
\begin{equation}
\text{Cov}(C_{\ell_1}, C_{\ell_2}) = \frac{1}{N-1} \sum_{i=1}^{N} \left( C_{\ell_1}^i - \bar{C}_{\ell_1} \right) \left( C_{\ell_2}^i - \bar{C}_{\ell_2} \right),
\end{equation}
where $N=1000$ is the number of mocks. We apply the Hartlap factor \citep{hartlap2007} $\frac{N-1}{N-p-1}$ to ensure the inverse covariance matrix is an unbiased estimator, where $p$ is the number of $C_\ell$s.

We present the correlation matrix normalised from the covariance matrix in 
Fig.~\ref{fig:corr4zbin} for four redshift bins and Fig.~\ref{fig:corrallzbin} for all redshift bins combined. The faint diagonal terms in the top-right and bottom-left panels of the correlation matrix plots represent correlations between \text{eBOSS} $\times$ \text{eBOSS} and \text{eBOSS} $\times$ \text{LoTSS} angular power spectra within the same redshift bins. These correlations arise from shared underlying large-scale structures within the same redshift range, while correlations between different redshift bins are weaker due to limited spatial and redshift overlap. Furthermore, since the plots are presented on a logarithmic scale, these faint correlations may appear visually stronger than their actual numerical values. Our results are within 1\% consistent with the Gaussian covariance at $\ell>50$, which is within our $\ell$ range applied in this paper; also, $\ell>50$ can avoid the deviation when we use the Limber approximation in \texttt{pyccl} compared with \texttt{CAMB}.

\subsection{Model comparison}

For the BAO template described in Eq.~(\ref{bao_model}), we tested different parameterisations listed in Table \ref{tab:model_comparison}. In the likelihood function, we used the average $C_\ell$ values obtained from 1000 simulated datasets (mocks) as the observed data points and the covariance matrix calculated from these mocks to account for uncertainties. The theoretical $C_\ell$ values from our model were used as the input to generate these mocks.

In the model for cross-correlation, we can neglect the shot noise term since we assume the shot noise between the two surveys is not correlated. However, we still need to add a number of polynomial terms to marginalise over the power spectrum broadband shape, leaving only the BAO wiggles as useful information to be fitted, while the other parameters are treated as nuisance parameters. To test whether our fit-parameter is non-Gaussian, we applied the $D$-statistics in the Kolmogorov-Smirnov (KS) test, which is defined as the maximum absolute difference between the empirical cumulative distribution function (ECDF) of the sample and the cumulative distribution function (CDF) of the reference distribution (in our case, the Gaussian distribution). We found that $D > 0.7$ for all tested parameterisations, indicating that due to the current S/N of the data, the $\alpha$ parameter is non-Gaussian. Thus we need to be cautious when interpreting the results, as many traditional methods of uncertainty estimation assume Gaussian distributions.

Based on the principles of a minimum variance unbiased estimator (MVUE), we define our model selection criteria as follows: the most appropriate model is one where (i) the mean value of $\alpha$ from the mock MCMC estimation is closest to 1, ensuring the estimator is unbiased, as $\alpha = 1$ is the expected value when the input and observed cosmological parameters match; and (ii) the 68\% confidence interval for $\alpha$ is within the prior range $[0.8, 1.2]$ with minimal error, indicating that the estimator has low variance and high precision.

As presented in Table \ref{tab:model_comparison}, the values of $\alpha$ are all within a percent level. For cross-correlation, we select the model $C_{\ell, x} = \left(\frac{B_x}{\alpha^2}\right) C_\ell^\mathrm{BAO} + a_{1,x} + \frac{a_{2,x}}{\ell} + a_{3,x} \ell$, which gives the a mean value of $\alpha = 1.0042$, where the bias of $\alpha$ is within 0.6\%, indicating sub-percent level accuracy. The 68\% confidence interval is $(0.9, 1.12)$, which falls within the prior range $(0.8, 1.2)$. For auto-correlation of eBOSS, the same type of model is also selected. Our template model is also in agreement with one of the models adopted in the Dark Energy Survey (DES; \citealp{des1, des2}) BAO analysis using the angular power spectrum, in which the template has been validated by both data and simulation.

\section{Results}\label{sec:results}

\subsection{Angular power spectrum}

With the validation by the mock catalogue presented above, we applied our chosen model and the pipeline used in the mock catalogue to real survey data. Figures~\ref{fig:cl-auto-mock} and \ref{fig:cl-cross-mock} present the auto- and cross-angular power spectra we measured, compared to the mean of the mock catalogues described in Sect. \ref{sec:choose_bz_nz}. For all redshift bins, the measurements are in good agreement. The error bars are derived from the covariance matrix computed from the 1000 mock catalogues for each survey.

We have detected significant cross-correlations in the $C_\ell$ measurements with significances of 7.82$\,\sigma$, 7.99$\,\sigma$, 8.01$\,\sigma$, and 9.16$\,\sigma$ for the respective redshift bins. These significances were derived using a frequentist likelihood ratio test, where the null hypothesis assumes no cross-correlation between LoTSS and eBOSS. When combining all redshifts, the significance increases to 14.73$\,\sigma$. These results indicate a substantial cross-correlation of galaxy sources between LoTSS and eBOSS, confirming our expectations described in Sect. \ref{sec:Intro}. Therefore, the cross-angular power spectrum can be effectively utilised for large-scale structure analysis, as detailed in the following section.

\subsection{BAO constraints}

We performed the BAO analysis in the following three ways: (i) with LoTSS $\times$ eBOSS cross-correlation in one redshift bin $0.6<z<0.66$ only, which contains most of the galaxies in eBOSS and secures a sufficiently thin redshift slice such that the BAO signal is not smoothed out by projecting along the radial direction; (ii) a combination of LoTSS $\times$ eBOSS cross-correlation and eBOSS $\times$ eBOSS auto correlation in the same redshift bin $0.6<z<0.66$, in order to see whether adding in the eBOSS dataset, leads to better constraints on the parameter. The covariance between auto- and cross-correlation has been accounted for, as shown in Fig.~\ref{fig:corr4zbin}; (iii) a combination of LoTSS $\times$ eBOSS cross-correlation in four redshift bins, assuming these redshifts share the same BAO ``shift'' parameter $\alpha$, which is viable when the redshift interval is relatively small (overall $\Delta z = 0.4$ in our case).

\begin{figure}
   \centering
   \includegraphics[width=\hsize]{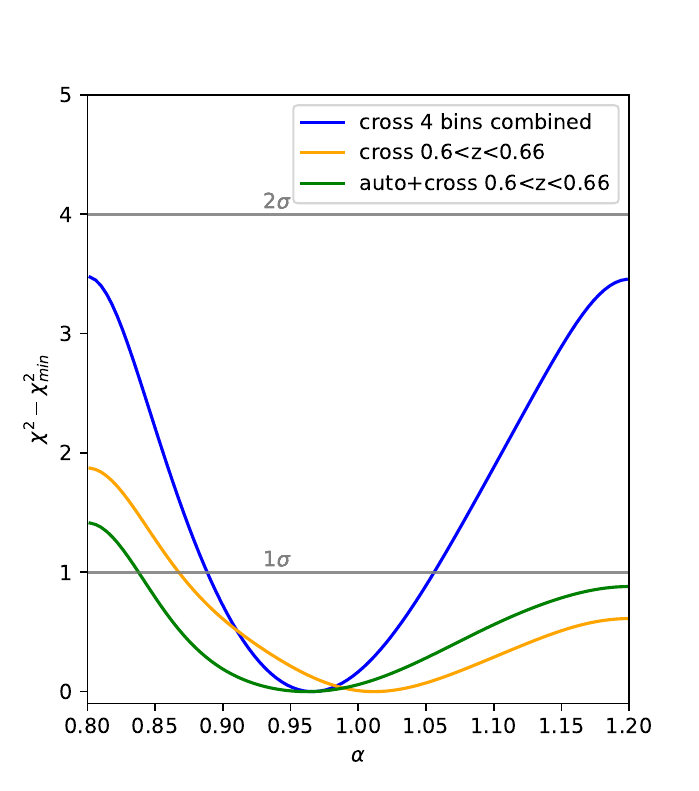}
      \caption{$\Delta \chi^2 \equiv \chi^2-\chi^2_{min}$ of the BAO constraints. The orange curve shows the BAO parameter $\alpha$ that is MCMC fitted using the cross angular power spectrum between LoTSS and eBOSS within one eBOSS redshift bin $0.6<z<0.66$; the blue curve shows the $\alpha$ parameter combining four eBOSS redshift bins in the cross angular power spectrum. The green curve shows the eBOSS auto angular power spectrum combined with cross angular power spectrum in the first redshift bin $0.6<z<0.66$. The grey lines at 1$\sigma$ and 2$\sigma$ correspond to the 68\% and 95\% confidence intervals, respectively, under the assumption of a Gaussian distribution for the parameter.}
      \label{fig:dchi2}
\end{figure}

\begin{figure}
   \centering
   \includegraphics[width=\hsize]{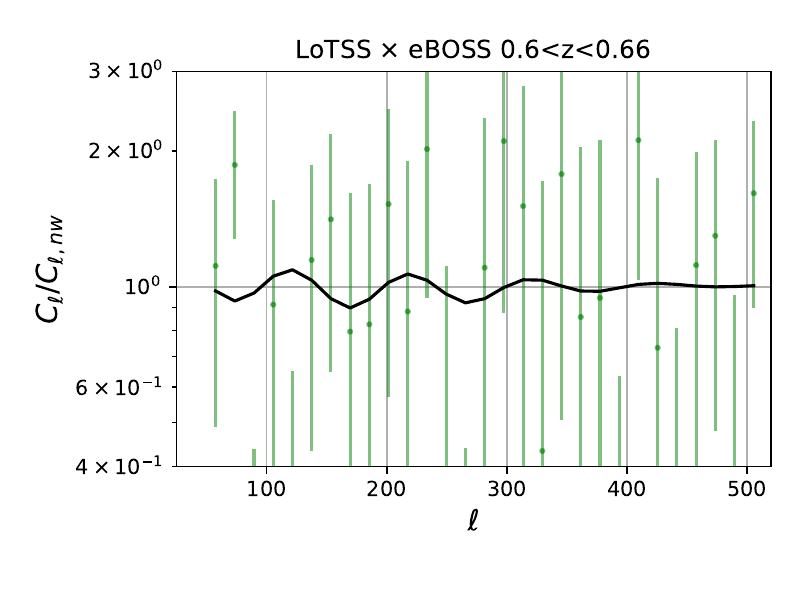}
      \caption{Best-fit cross angular power spectrum in the first redshift bin: $0.60<z<0.66$, data measurements (green) compared with the fitting results (black), normalised with the no-wiggle $C_\ell$s.}
      \label{fig:bestfit_cross_z1}
\end{figure}

\begin{figure}
   \centering
   \includegraphics[width=\hsize]{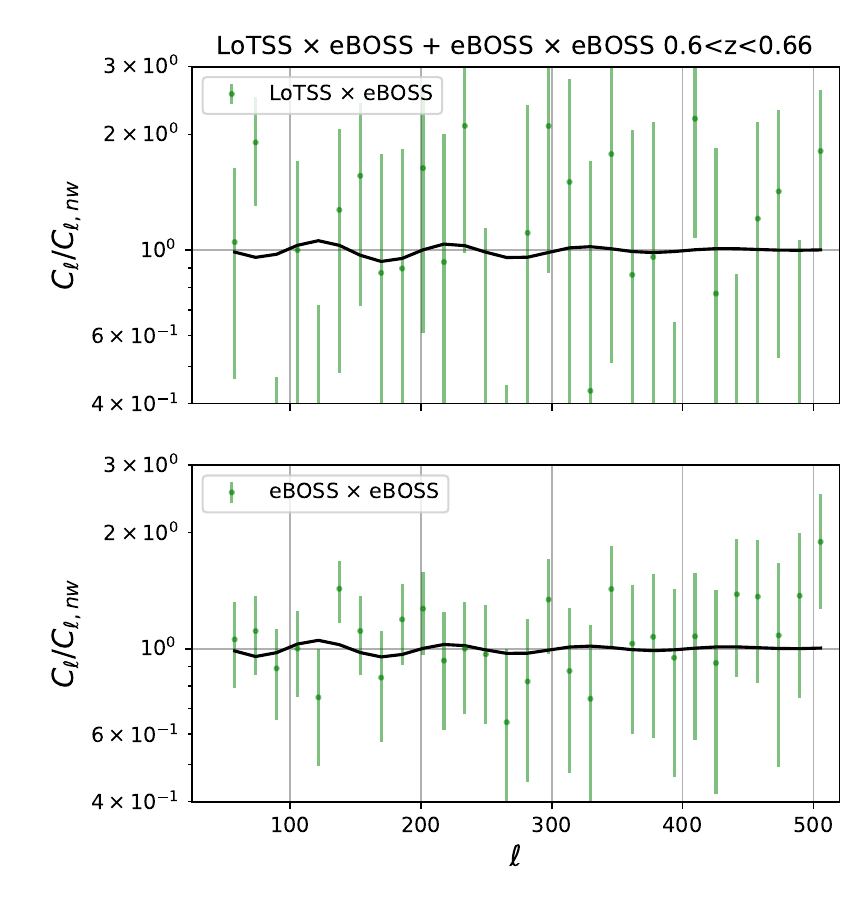}
      \caption{Best-fit LoTSS $\times$ eBOSS + eBOSS auto angular power spectrum in the first redshift bin: $0.60<z<0.66$, data measurements (green) compared with the fitting results (black), normalised with the no-wiggle $C_\ell$s.}
      \label{fig:bestfit_autocross_z1}
\end{figure}

\begin{figure}
   \centering
   \includegraphics[width=\hsize]{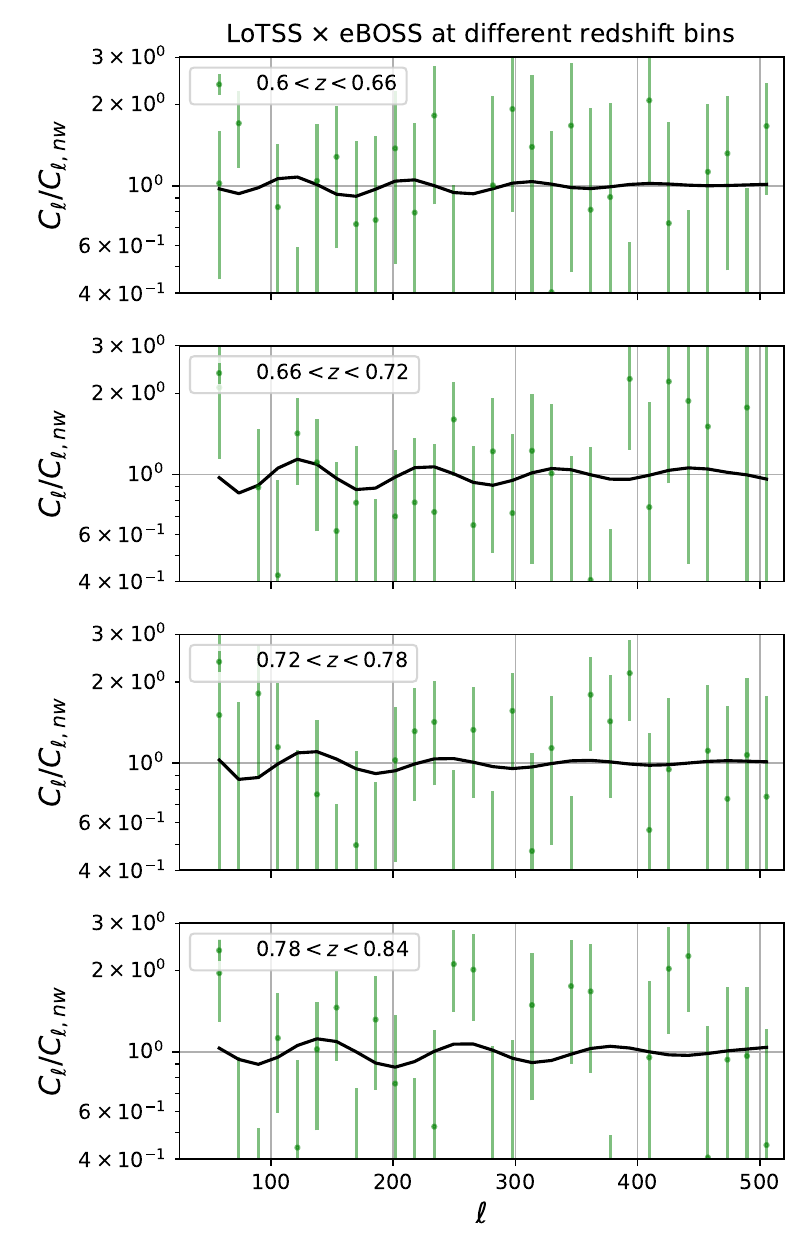}
      \caption{ Best-fit LoTSS $\times$ eBOSS cross angular power spectrum combining four redshift bins, data measurements (green) compared with the fitting results (black), normalised with the no-wiggle $C_\ell$s.}
      \label{fig:bestfit_cross_4bins}
\end{figure}

We report evidence of a BAO signal from the cross-correlation between LoTSS DR2 and eBOSS LRGs ($0.6<z<0.66$) shown in Table~\ref{table:main_res} and in Fig.~\ref{fig:dchi2} for all cases, along with the best-fit $C_\ell$s compared to data is presented in Fig.~\ref{fig:bestfit_cross_z1} for case (i), Fig.~\ref{fig:bestfit_autocross_z1} for case (ii), and Fig.~\ref{fig:bestfit_cross_4bins} for case (iii). For all parameters, we set flat priors $\alpha \in [0.8,1.2]$, $B_0 \in [0,10]$, $a_1,a_2,a_3 \in [-100,100]$, respectively, which are consistent with the settings in the mock test. We used \texttt{Cobaya} \citep{torrado2021cobaya} for Markov chain Monte Carlo (MCMC) parameter estimation, and marginalise over the nuisance parameters $b$, $a_1$, $a_2$, and $a_3$. In particular, for one redshift bin in cross-correlation, we obtained the BAO shift parameter $\alpha=1.01 \pm 0.11$. The reduced chi-squared value is $0.85$, calculated as the ratio of the minimum chi-squared to the degrees of freedom (dof). There is a 1\% deviation in the mean of $\alpha$ from the fiducial value. As shown in Fig.~\ref{fig:dchi2}, the cross-correlation errors are large, and the BAO wiggles in the angular power spectrum are relatively weak. This situation can lead to slight overfitting of the parameter, where some noise fluctuations in the cross-correlation measurements might be misinterpreted as signal. Given that our default fitting model is nonlinear, the reduced $\chi^2$ statistic may not fully capture the goodness of fit \citep{reducedchi2}.

\bgroup
\def\arraystretch{1.5}
\begin{table*} 
    \centering
    \caption{BAO constraints from the LoTSS–eBOSS cross-correlation and its combination with the eBOSS auto-correlation.}
    \label{table:main_res}
    \renewcommand{\arraystretch}{1.5}
    \begin{tabular}{lccccc}
    \toprule
    Measurement & $z_{\rm eff}$ & $\alpha$ & $D_A(z_{\rm eff})/r_\mathrm{d}$ & $\chi^2_\mathrm{min}/\mathrm{dof}$ & $\Delta\chi^2_\mathrm{nw}$ \\ 
    \midrule
    Cross ($0.6<z<0.66$) & 0.63 & $1.01\pm 0.11$ & $9.997\pm 1.089$ & 21.20/25 & 5.02 \\  
    Cross + auto ($0.6<z<0.66$) & 0.63 & $0.988^{+0.080}_{-0.160}$ & $9.779^{+0.792}_{-1.584}$ & 30.66/50 & 7.27 \\  
    Cross (4 bins combined) & 0.72 & $0.968^{+0.060}_{-0.095}$ & $10.122^{+0.627}_{-0.993}$ & 99.72/103 & 16.28 \\  
    \bottomrule
        \end{tabular}
    \tablefoot{The $\alpha$ parameter is converted to angular diameter distance $D_A$ using Eq.~\ref{eq: alpha-def}, with a fiducial cosmology from Planck 2018.  
    $\chi^2_\mathrm{min}/\mathrm{dof}$ denotes the reduced chi-squared of our fits. Also,  
    $\Delta\chi^2_\mathrm{nw}$ compares our results with models lacking BAO features ('nw' model; see \citealt{eihu1998}).}
    
\end{table*}
\egroup

\begin{figure}
    \centering
    \includegraphics[width=1\hsize]{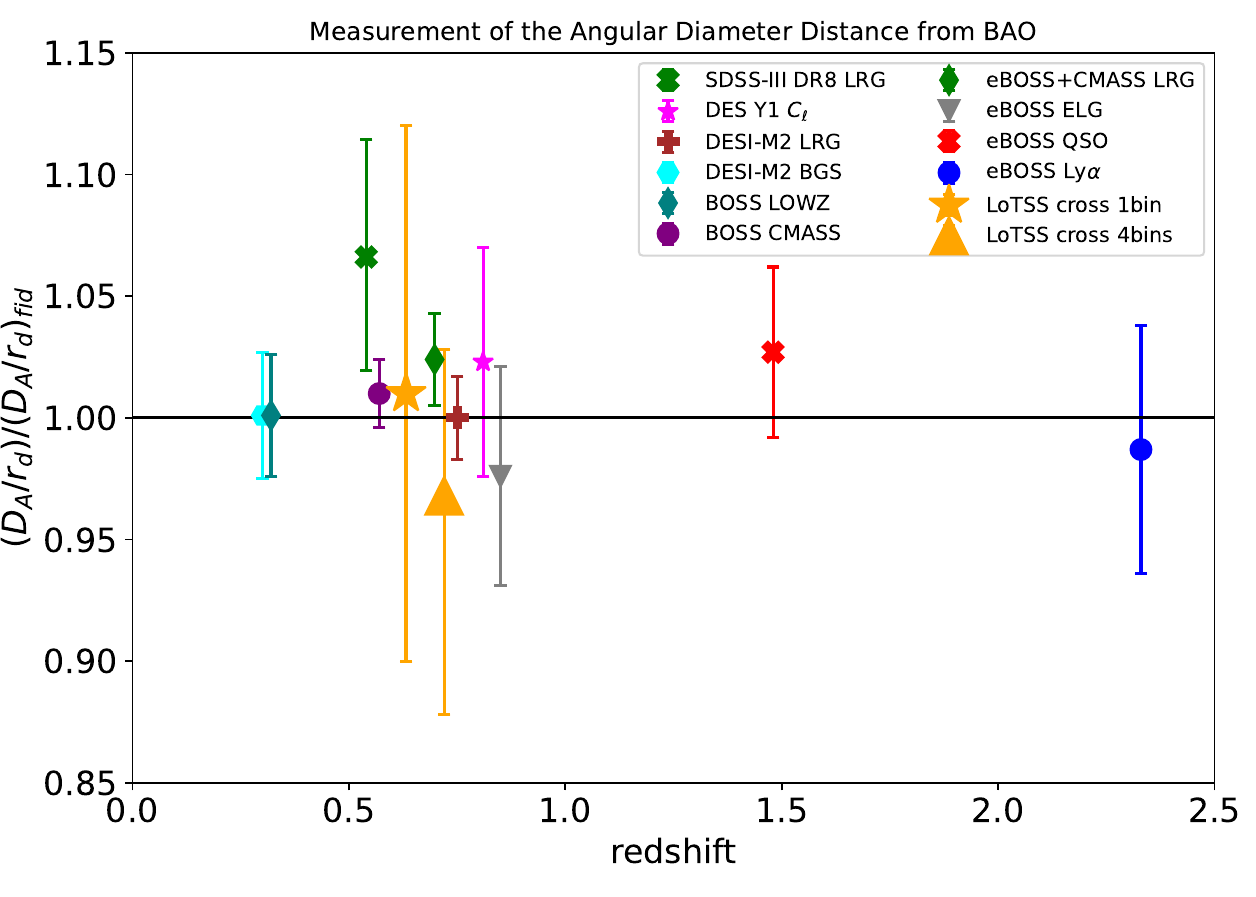}
    \caption{ BAO dilation parameter $\alpha$ i.e. the ratio of $D_A/r_d$ to fiducial ones measured in this work (orange) compared to previous works. Presented are LoTSS $\times$ eBOSS in redshift bin $0.6<z<0.66$ (orange star), LoTSS $\times$ eBOSS combining all redshift bins (orange triangular). Other measurements are labelled and described in the figure legend and main text. Note: for measurements using 3D power spectrum, we adopt the $\alpha_{\perp}$ results; for  works using the angular power spectrum, e.g. SDSS-DR8 and DES-Y1, we adopt the $\alpha$ parameter measurements.}
    \label{fig:res_comparison}
\end{figure}

With the combination of eBOSS auto $C_\ell$ values at the same redshift bin, also taking the cross covariance between eBOSS auto and cross $C_\ell$ values shown in the large correlation matrix in Fig. \ref{fig:corr4zbin} into account, we report an $\alpha$ parameter with smaller errors but higher bias, shown in Table \ref{table:main_res}. This is likely due to the introduction of three extra nuisance parameters ($B_e, a_{1,e}, a_{2,e},a_{3,e}$). While these parameters help in reducing errors by better fitting the eBOSS auto correlation measurements, they can also lead to potentially overfitting.

Finally, by combining the cross-correlation $C_\ell$s in four redshift bins, with an effective redshift of $z=0.72$, we obtained our best fit $\alpha$ measurement with the tightest constraints by making the assumption that these redshift bins share the same BAO parameter. We used the 1$\sigma$ and 2$\sigma$ horizontal line as references (shown in Fig.~\ref{fig:dchi2}) although our parameter is non-Gaussian-distributed and the intersection might not fully represent the real 68\% and 95\% confidence interval in the fitting. We show that by combining four redshift bins, the errors are reduced by 2\%, while the parameter distribution exhibits increased symmetry and reduced tails that align more closely with a Gaussian distribution. We also have included a contour plot of $\alpha$ and the shared nuisance parameter $a_{1,x,z1} $in Fig. \ref{fig:corner} for the three cases.

To assess the statistical significance of BAO detection, we perform a null hypothesis test, where the null hypothesis ($H_0$) assumes that the observed angular power spectrum follows a no-wiggle template with no BAO features. The alternative hypothesis ($H_1$) assumes the presence of BAO wiggles. The statistical preference for the BAO model was quantified using the $\Delta\chi^2_{\text{nw}}$ metric:

\begin{equation}
\Delta\chi^2_{\text{nw}} = \chi^2_{\text{no-wiggle}} - \chi^2_{\text{BAO}}.
\end{equation}

The quantity \(\chi^2_{\text{no-wiggle}}\) represents the chi-squared value obtained from fitting the data with a no-wiggle template; that is, a model that captures the broadband shape of the angular power spectrum without including BAO features. This no-wiggle template is defined as described in \citealp{eihu1998}. In contrast, \(\chi^2_{\text{BAO}}\) is the chi-squared value for the best-fit model that incorporates BAO wiggles. Therefore, the difference \(\Delta\chi^2_{\text{nw}}\) quantifies the improvement in the fit when BAO features are included. A larger \(\Delta\chi^2_{\text{nw}}\) indicates a stronger statistical preference for the presence of BAO wiggles relative to the no-BAO (null) model. Assuming that \(\Delta\chi^2_{\text{nw}}\) approximately follows a chi-squared distribution with one degree of freedom, we computed the corresponding p-values for our different cases. For the single redshift bin, a \(\Delta\chi^2_{\text{nw}} = 5.02\) yields a p-value of about 2.5\%, meaning that if the no-BAO model were correct, there is only a 2.5\% chance of obtaining such a chi-squared improvement purely by random fluctuations. In the case where we combined the cross-correlations with eBOSS auto-correlations in the same redshift bin, we find that \(\Delta\chi^2_{\text{nw}} = 7.27\) corresponds to a p-value of roughly 0.7\%; whereas when combining four redshift bins, \(\Delta\chi^2_{\text{nw}} = 16.28\) results in a p-value of approximately 0.01\%. These low p-values indicate that the null hypothesis (i.e. the absence of BAO wiggles) is strongly disfavoured.

Since the distribution of the BAO parameter $\alpha$ is non-Gaussian, reporting detection significance in sigma levels may not be appropriate. Therefore, we focus here on the $\Delta \chi^2$ values and provide 68\% confidence intervals for $\alpha$ in Table~\ref{table:main_res}, which avoid assumptions of Gaussianity and offer a statistically robust interpretation of our results. 

To further validate the robustness of our results, we performed a jackknife resampling analysis for the eBOSS $\times$ LoTSS cross-correlation in the first redshift bin $0.6<z<0.66$. Specifically, we generated 1000 jackknife realisations by dividing the survey footprint into 1000 equal-area subregions and applied a leave-one-out method to compute the cross-angular power spectrum. Additionally, we conducted a ten-subregion jackknife test using a similar leave-one-out procedure over ten equal-area subregions. The BAO parameter $\alpha$ was then measured using the covariance matrix derived from both the 1000-subregion jackknife and 10-subregion jackknife samples. The results from the 1000-subregion jackknife test yield $\alpha = 1.0087 \pm 0.1067$ with a reduced $\chi^2 = 0.82$, which is consistent with the full dataset measurement. The jackknife-derived error bars are slightly smaller than those from the full dataset, as expected, due to the reduction in cosmic variance in jackknife resampling. The mean $\alpha$ estimates remain stable within the range [0.95, 1.05] and the reduced $\chi^2$ values for these fits range from 0.8 to 1.4, further confirming the reliability of our measurements. We present the full set of jackknife results, including the distribution of $\alpha$ values and their dependence on the reduced $\chi^2$, in Fig. \ref{fig:jackknife}.

In comparison to previous works on BAO detection, we present the $\alpha$ parameter, specifically $\alpha_{\perp}$, which indicates the angular BAO distance ladder in our results and other surveys, as shown in Fig.~\ref{fig:res_comparison}. We compare our findings with those from previous studies that utilised the angular power spectrum, such as DES~\citep{des1} and SDSS-DR8~\citep{seo2008nonlinear}. Additionally, we include comparisons with 3D BAO analyses, including BOSS DR12 LOWZ+CMASS \citep{gil2016clustering}, eBOSS LRG \citep{bautista2018sdss}, eBOSS ELG \citep{ebosselg-tamone2020completed}, eBOSS QSO \citep{hou2021completed}, and eBOSS Lyman-$\alpha$ \citep{lyalpha}. We also made a comparison with the most recent DESI BAO detections \citep{desibao}. Our results are consistent with the fiducial model within the 68\% confidence level, although our result exhibits errors that are two to three times larger, compared to previous studies that have higher precision of redshift information and larger number of sources. The larger errors arise from the non-Gaussian distribution of our $\alpha$ measurements in the LoTSS data, in contrast to the Gaussian distributions typically found in optical surveys. These surveys benefit from stronger S/Ns, allowing them to achieve smaller error bars and more precise measurements of $\alpha$. As a result, their $\alpha$ distributions are symmetric with smaller 68\% confidence interval ranges.

\subsection{Number density weighted bias and the bias constraints}

Alternatively, we could make use of the eBOSS spectroscopic redshifts to infer the bias and redshift distribution of LoTSS radio sources. We used the template described in 
Eqs.~(\ref{eq:pk2cl}) and (\ref{eq:WAB}) by fixing the bias and redshift distribution of eBOSS, $b_A$ and $p_A$, as well as the redshift distribution of LoTSS, $p_B$, such that we are able to measure the bias of LoTSS $b_B$, assuming either a constant bias, $b_0$, or a bias form, $b_D/D(z)$, that evolves inversely proportional to the linear growth function. We measured a flat bias of $b = 2.64 \pm 0.20$ and an evolving bias of $b_D = 1.80 \pm 0.13$ at  redshifts of $0.6<z<1$. We also measured the biases in each redshift bin, where we obtained $b_C=2.05\pm0.33$ for $0.6<z<0.66$, $b_C=2.40 \pm 0.36$  for $0.66<z<0.72$, $b_C=2.69 \pm 0.42$ for $0.72<z<0.78$, and $b_C=3.11\pm 0.46$ for $0.78<z<0.84$.  We did not incorporate higher redshift bins due to much fewer galaxies in the eBOSS catalogue and only a very slight cross-correlation in the $C_\ell$  leading to a large measured bias and variance.

\begin{figure}
    \centering
    \includegraphics[width=1.1\linewidth]{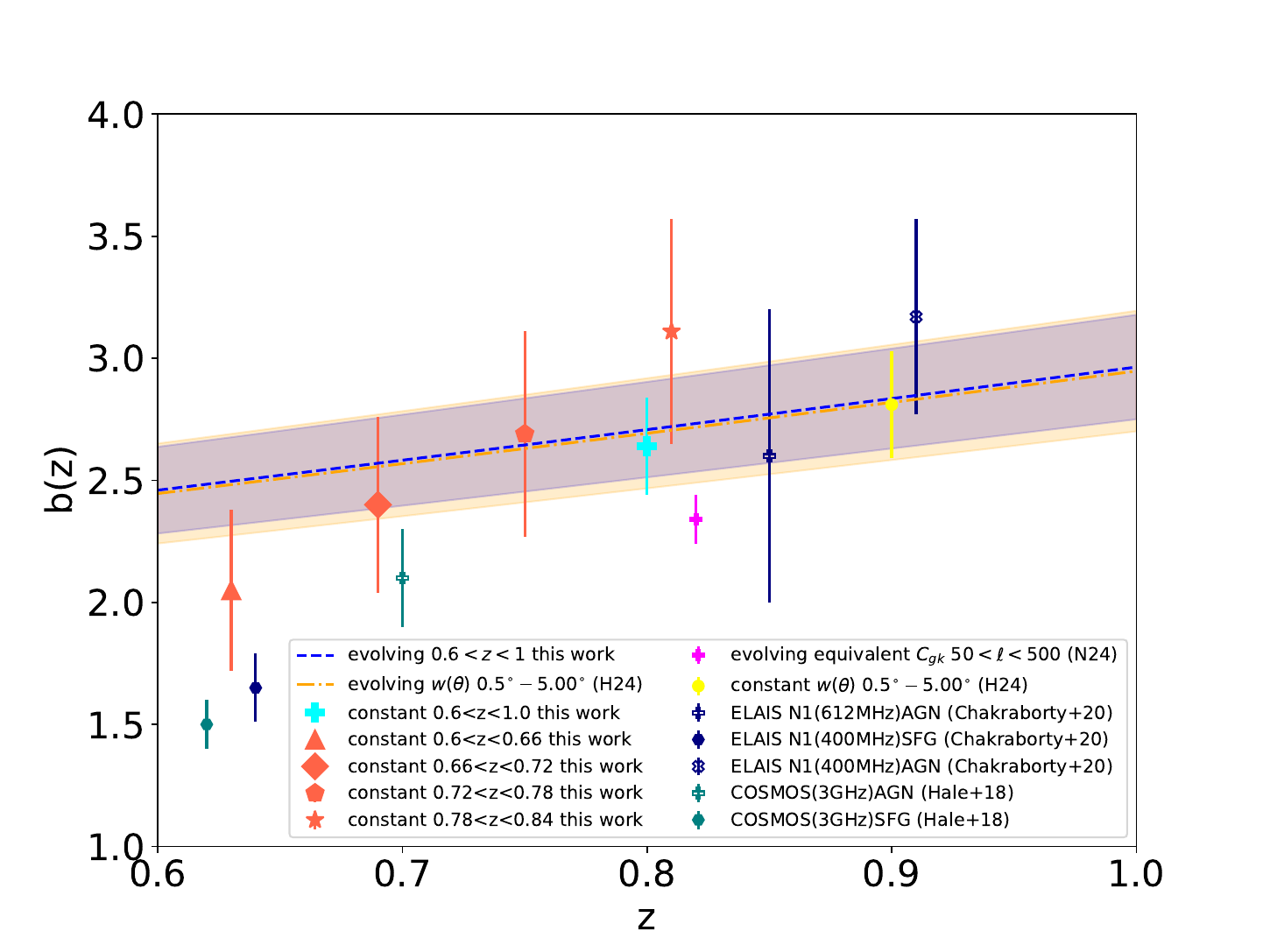}
    \caption{
Bias values measured in this work and compared with previous studies. The evolving bias model derived from this work ($0.6<z<1$) is shown as a dashed blue line with a shaded error band, while the evolving bias model from H24 is shown as a dot-dashed orange line with a shaded band. The constant bias over the full redshift range ($0.6<z<1$) from this work is shown as a cyan pentagon. Constant biases in four redshift bins from this work are shown as Tomato-colored markers with different shapes. Additional comparisons include the evolving bias derived from LoTSS $\times$ CMB lensing cross-correlation (N24), plotted as magenta diamonds; the constant bias from H24 is shown as a yellow circle. The AGN and SFG biases from \citet{hale2018clustering} and \citet{chakraborty2020study} are plotted as hollow markers with navy and teal edges, respectively, reflecting their frequency bands and galaxy types. All error bars represent $1\sigma$ uncertainties.}
    \label{fig:bias}
\end{figure}

A comparison between the bias in this work and previous works is presented in Fig.~\ref{fig:bias}. As seen in the figure, the evolving bias across all redshifts in our work shows a significant agreement with the bias obtained in H24 in real space for the scale range $36 < \ell < 3600$. It is also within 1$\sigma$ error when compared to the equivalent evolving bias measured in CMB lensing cross-correlating with LoTSS DR2 for the same scale range ($50 < \ell < 500$) and using the same cosmology model (N24). The bias values measured in the four separate redshift bins show a slightly different trend compared to the overall evolving bias model. This may be due to the fact that, while the first three bins have reasonable reduced $\chi^2$ values close to 1, the fourth bin has a larger reduced $\chi^2$ value of 1.33, indicating a relative poorer fitting. Also, it is worth noting that our bias is in general slightly larger than the works that measure the SFG bias \citep{hale2018clustering,chakraborty2020study}. We believe this is due to the fact that we are cross-correlating LRGs observed in an optical survey with radio sources that consist of both SFGs and active galactic nuclei (AGNs), while LRGs have typically low star-forming rates and some of them are found to host AGNs. As observed in the LOFAR Deep Fields DR1, a transition occurs at approximately 1.5 mJy at 150 MHz, where the population shifts from being dominated by SFGs to being dominated by radio-loud AGNs \citep{best2023lofar}. This shift indicates that above the 1.5 mJy flux limit adopted in this work, most radio galaxies are AGNs (see also Fig. 9 in \citealp{best2023lofar}). As a result, the sources in our radio catalog selected by cross-correlation are more likely AGNs, which are typically more strongly clustered and generally have a higher bias than SFGs \citep{agn1,agn2,agn3,agn4,agn5}. Thus, the bias by averaging is larger than in works based on the assumption that galaxy types are subject to a similar population proportion.

\section{Discussion and conclusion} \label{sec:conclusion}

In this work, we have calculated the cross-correlation between LoTSS DR2 and eBOSS LRGs in harmonic space, utilising the overlapping footprints between both surveys. We constructed a mask that accounts for systematics in LoTSS and in eBOSS. Specifically, by constructing a correlated mock galaxy catalogue for both LoTSS and eBOSS, we validated our pipeline, obtained the covariance matrix of the angular power spectra, selected the appropriate BAO model, and applied them to real survey data. In the BAO template, we also accounted for the solid angle conversion in harmonic space. With the data and theoretical angular power spectra, we were able to measure the BAO dilation parameter $\alpha$, which is a ratio of the measured BAO scale compared to a fiducial value adopted from the best-fit cosmological model of Planck. To ensure the robustness of our measurement, we performed a jackknife resampling test by computing 10 and 1000 jackknife realisations, which showed stable $\alpha$ measurements and slightly smaller error bars due to reduced cosmic variance. We also conducted a null hypothesis test using the no-wiggle BAO power spectrum, confirming the statistical preference for BAO features. For the very first time, we see evidence for BAO in the cross-correlation of a radio continuum and an optical survey, which provides an important consistency check of our current understanding of cosmology. On the other hand, we measured the linear bias values of radio sources in LoTSS as a function of redshift, which are in good agreement with the ones calculated in configuration space.
Despite the statistically significant cross-correlation in the $C_\ell$s and the first evidence for a BAO signal seen in the radio continuum, there are still a number of limitations that need to be addressed: 

\begin{itemize}
    \item Although the LoTSS DR2 survey covers a wide redshift range, our cross-correlation is limited to the range available in the eBOSS survey because precise redshift information is only available from this spectroscopic survey. The projection along the line of sight smooths out BAO signals, requiring us to use thin redshift bins, which results in a loss of galaxy information from wider bins. Additionally, compared to the most up-to-date optical surveys, the limited overlap in survey areas restricts the precision of our measurements.
    \item The mock catalogue we generated here assumes a lognormal distribution for the overdensity map, which is sufficient for an angular correlation. However, more accurate mocks that mimic the radio catalogue distribution and account for systematics could improve future studies. Currently, we lack the knowledge to produce N-body simulations, as typically done in optical surveys, due to insufficient information for modelling black holes in AGNs. This makes it difficult to simulate sub-grid cosmological gravity.
    \item The distribution of the dilation parameter $\alpha$ we measure tends to be non-Gaussian, making it difficult to perform a complete model selection test. The non-Gaussian distribution of the $\alpha$ parameter requires the development of additional statistical tools to quantify such cases, especially when the signal is relatively weak and exhibits non-Gaussian properties, such as a multi-modal distribution.
    
\end{itemize}

Our study demonstrates the feasibility of conducting BAO analysis using radio continuum data. This approach shows significant promise for detecting a strong BAO signal in future observations. We outline several possibilities for future research:

\begin{itemize}
\item With the aid of spectroscopic redshifts from WEAVE-LOFAR \citep{smith2016weave}, which has a high redshift success rate approaching 100 percent at $z < 1$, we will be able to better understand the nature of the radio sources and to construct a 3D overdensity map. These steps will enable a 3D BAO analysis.

\item Up-to-date spectroscopic surveys such as Dark Energy Spectroscopic Instrument (DESI; \cite{aghamousa2016desi,desibao}) and Euclid  space telescope \citep{euclidforecast} will, on the other hand,  also provide a more complete overlapping region among cosmological surveys in different bands. Thus, it could be possible to detect a more significant BAO signal in cross-crrelations with large radio surveys such as LoTSS. 

\item Moreover, the cross-correlation in configuration space and higher order statistics will help amend the power spectrum estimations in Fourier space. 

\end{itemize}

The forthcoming completion of the LoTSS, which will cover 80\% of the northern hemisphere, presents an exciting opportunity to enhance BAO signal detection. This can be achieved through either 2D analysis via cross-correlation or 3D analysis incorporating redshift information. The survey's deep reach of redshift, offers a unique prospect for testing models of dark energy and gravity. This expanded dataset will allow for more precise measurements and provide a robust framework for exploring the evolution of the Universe.

\begin{acknowledgements} 

We thank eBOSS collaboration and David Alonso, Julian Bautista, Cheng Zhao, Minas Karamanis for insightful discussions. 

JZ acknowledge support by the project "NRW-Cluster for data intensive radio astronomy:
Big Bang to Big Data (B3D) “funded through the programme "Profilbildung 2020", an
initiative of the Ministry of Culture and Science of the State of North Rhine-Westphalia.“ PT acknowledges the support of the RFIS grant (No. 12150410322) by the National Natural Science Foundation of China (NSFC). GBZ is supported by the National Key R \& D Program of China (2023YFA1607800, 2023YFA1607803), NSFC grants 11925303 and 11890691, by the CAS Project for Young Scientists in Basic Research (No. YSBR-092), by science research grants from the China Manned Space Project with No. CMS-CSST-2021-B01, and the New Cornerstone Science Foundation through the XPLORER prize.
DJS acknowledges support from the Bundesministerium f\"ur Bildung und Forschung (BMBF) ErUM-Pro grant 05A20PB1 and Ministerium f\"ur Kultur und Wissenschaft des Landes Nordrhein-Westfahlen Profilbildung 2020 grant B3D.
CH’s work is funded by the Volkswagen Foundation. CH acknowledges additional support from the
Deutsche Forschungsgemeinschaft (DFG, German Research Foundation) under Germany’s Excellence Strategy EXC 2181/1 - 390900948
(the Heidelberg STRUCTURES Excellence Cluster).
SC acknowledges support from the Italian Ministry of University and Research, PRIN 2022 `EXSKALIBUR -- Euclid-Cross-SKA: Likelihood Inference Building for Universe Research', from the Italian Ministry of Foreign Affairs and International
Cooperation (grant no.\ ZA23GR03), and from the European Union -- Next Generation EU.
SJN is supported by the US National Science Foundation (NSF) through grant AST-2108402. MP acknowledges support from the Bundesministerium für Bildung und Forschung (BMBF) ErUM-IFT 05D23PB1.
LOFAR is the Low Frequency Array designed and constructed by ASTRON. It has observing, data processing, and data storage facilities in several countries, which are owned by various parties (each with their own funding sources), and which are collectively operated by the ILT foundation under a joint scientific policy. The ILT resources have benefited from the following recent major funding sources: CNRS-INSU, Observatoire de Paris and Université d’Orléans, France; BMBF, MIWF-NRW, MPG, Germany; Science Foundation Ireland (SFI), Department of Business, Enterprise and Innovation (DBEI), Ireland; NWO, The Netherlands; The Science and Technology Facilities Council, UK; Ministry of Science and Higher Education, Poland; The Istituto Nazionale di Astrofisica (INAF), Italy.

Funding for the Sloan Digital Sky Survey V has been provided by the Alfred P. Sloan Foundation, the Heising-Simons Foundation, the National Science Foundation, and the Participating Institutions. SDSS acknowledges support and resources from the Center for High-Performance Computing at the University of Utah. SDSS telescopes are located at Apache Point Observatory, funded by the Astrophysical Research Consortium and operated by New Mexico State University, and at Las Campanas Observatory, operated by the Carnegie Institution for Science. The SDSS web site is \url{www.sdss.org}.

SDSS is managed by the Astrophysical Research Consortium for the Participating Institutions of the SDSS Collaboration, including Caltech, The Carnegie Institution for Science, Chilean National Time Allocation Committee (CNTAC) ratified researchers, The Flatiron Institute, the Gotham Participation Group, Harvard University, Heidelberg University, The Johns Hopkins University, L’Ecole polytechnique f\'{e}d\'{e}rale de Lausanne (EPFL), Leibniz-Institut f\"{u}r Astrophysik Potsdam (AIP), Max-Planck-Institut f\"{u}r Astronomie (MPIA Heidelberg), Max-Planck-Institut f\"{u}r Extraterrestrische Physik (MPE), Nanjing University, National Astronomical Observatories of China (NAOC), New Mexico State University, The Ohio State University, Pennsylvania State University, Smithsonian Astrophysical Observatory, Space Telescope Science Institute (STScI), the Stellar Astrophysics Participation Group, Universidad Nacional Aut\'{o}noma de M\'{e}xico, University of Arizona, University of Colorado Boulder, University of Illinois at Urbana-Champaign, University of Toronto, University of Utah, University of Virginia, Yale University, and Yunnan University.

This research was conducted using Python 3 \citep{python3:2009} along with several key software packages that were crucial for our analysis, including:
healpy \citep{zonca2019healpy},
HEALPix \citep{gorski1999healpix},
Astropy \citep{astropy:2013, astropy:2018, astropy:2022},
pymaster \citep{alonso2019unified},
pyccl \citep{Chisari_2019},
getdist \citep{Lewis:2019},
NumPy \citep{NumPy:2020},
SciPy \citep{Virtanen:2020},
IPython \citep{ipython:2007},
Pandas \citep{pandas:2010},
Matplotlib \citep{Matplotlib:2007},
and seaborn \citep{seaborn:2021}.
\end{acknowledgements}

\bibliographystyle{aa}
\bibliography{aa51958-24}

\begin{appendix} 
\section{BAO model comparison using mock catalogue}\label{sec:apppen}

This section presents the BAO model comparison, using the mock catalogue for testing and how we choose the optimised model for the real survey data measurements. As presented in Table~\ref{tab:model_comparison}, for both the cross-correlation power spectrum and the eBOSS auto-correlation power spectrum, we opt for the model which is mostly unbiased; i.e. the mean value of the $\alpha$ parameter is closest to 1, while the 68\% confidence interval falls into the prior range and is the narrowest, since the observed BAO signal is relatively weak due to the power spectrum projection along the line of sight. Thus the final model (highlighted in the table) we chose is 
\begin{equation}
C_{\ell,i}  =\frac{B_i}{\alpha^2} C_\ell^\mathrm{BAO}+a_{1,i}+a_{2,i}/\ell+a_{3,i}\ell, 
\end{equation}
with the index $i \in \{\mathrm{e, x}\}$ either denoting the eBOSS auto-correlation or the LoTSS-eBOSS cross-correlation case.
\FloatBarrier
\bgroup
\def\arraystretch{1.5}
\begin{table}
\caption{Mock-based model comparison of BAO fitting in cross angular power spectrum and eBOSS auto angular power spectrum.}
\begin{tabular}{lll}
\toprule
model & $\alpha$ & 68\% CI      \\ \hline
$C_{\ell, x}  =\frac{B_x}{\alpha^2} C_\ell^\mathrm{BAO}$                                                                        & 1.0125   & (0.976,1.2)  \\
$C_{\ell, x}  =\frac{B_x}{\alpha^2} C_\ell^\mathrm{BAO}+a_{1,x}$                                                                & 1.0041   & (0.8,1.2)    \\
$C_{\ell, x}  =\frac{B_x}{\alpha^2} C_\ell^\mathrm{BAO}+a_{1,x}/\ell$                                                           & 1.0119   & (0.964,1.2)  \\
$C_{\ell, x}  =\frac{B_x}{\alpha^2} C_\ell^\mathrm{BAO}+a_{1,x}+a_{2,x}/\ell$                                                   & 1.0065   & (0.8,1.2)    \\
$C_{\ell, x}  =\frac{B_x}{\alpha^2} C_\ell^\mathrm{BAO}+a_{1,x}+a_{2,x}/\ell+a_{3,x}/\ell^2$                                    & 1.0204   & (0.8,1.2)    \\
$C_{\ell, x}  =\frac{B_x}{\alpha^2} C_\ell^\mathrm{BAO}+a_{1,x}+a_{2,x}/\ell+a_{3,x}\ell$                                       & 1.0042   & (0.9,1.12)   \\
$C_{\ell, x}  =\frac{B_x}{\alpha^2} C_\ell^\mathrm{BAO}+a_{1,x}+a_{2,x}/\ell^2+a_{3,x}\ell$                                     & 1.0089   & (0.8,1.2)    \\ \hline
$C_{\ell, e} = \frac{B_e}{\alpha^2} C_\ell^\mathrm{BAO}+a_{1,e}$                                                                       & 1.0134   & (0.982,1.2)  \\
$C_{\ell, e} = \frac{B_e}{\alpha^2} C_\ell^\mathrm{BAO}+a_{1,e}/\ell$                                                                  & 1.0305   & (0.967,1.2)  \\
$C_{\ell, e} = \frac{B_e}{\alpha^2} C_\ell^\mathrm{BAO}+a_{1,e}+a_{2,e}/\ell$                                                          & 1.0062   & (0.96, 1.19) \\
$C_{\ell, e} = \frac{B_e}{\alpha^2} C_\ell^\mathrm{BAO}+a_{1,e}+a_{2,e}/\ell+a_{3,e}/\ell^2$                                           & 1.0019   & (0.993,1.2)  \\
$C_{\ell, e} = \frac{B_e}{\alpha^2} C_\ell^\mathrm{BAO}+a_{1,e}+a_{2,e}/\ell+a_{3,e}\ell$                                              & 1.0085   & (0.92,1.13)  \\
$C_{\ell, e} = \frac{B_e}{\alpha^2} C_\ell^\mathrm{BAO}+a_{1,e}+a_{2,e}/\ell^2+a_{3,e}\ell$                                            & 1.0146   & (0.952,1.2)  \\
\bottomrule
\label{tab:model_comparison}
\end{tabular}
\end{table}
\egroup

\FloatBarrier

\section{Additional validation tests}\label{sec:apppen1}

In this appendix, we present additional figures supporting the robustness of our analysis.

\subsection{MCMC contour plots}
\begin{figure}
    \centering
    \includegraphics[width=0.8\hsize]{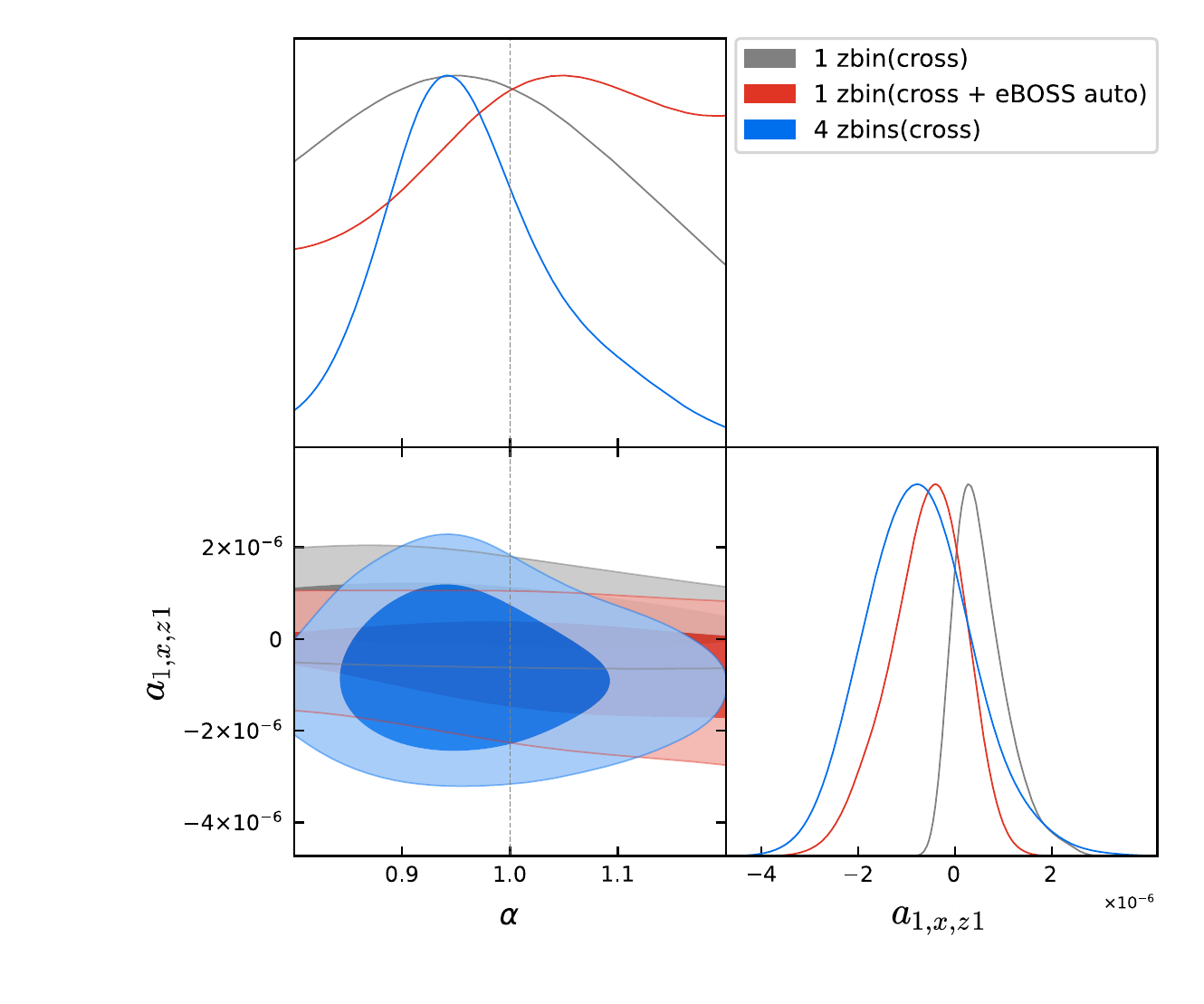}
    \caption{Contours of the parameter $\alpha$ and one of the nuisance parameters, $a_{1,x,z1}$, shown for all three cases, overlaid for comparison. The gray contours represent the LoTSS $\times$ eBOSS cross-correlation in a single redshift bin ($0.6<z<0.66$), while the red and blue contours correspond to the cases where the cross-correlation is combined with the eBOSS auto-correlation in the same redshift bin and where the cross-correlation is combined across four redshift bins, respectively.}
    \label{fig:corner}
\end{figure}

Figure~\ref{fig:corner} shows the MCMC posterior distributions for the BAO parameter $\alpha$ and a shared nuisance parameter, $a_{1,x,z1}$, in all three cases considered in this work. The different contour colors correspond to different dataset combinations, as described in the main text.

\subsection{Jackknife resampling results}
\begin{figure}
    \centering
    \includegraphics[width=1\hsize]{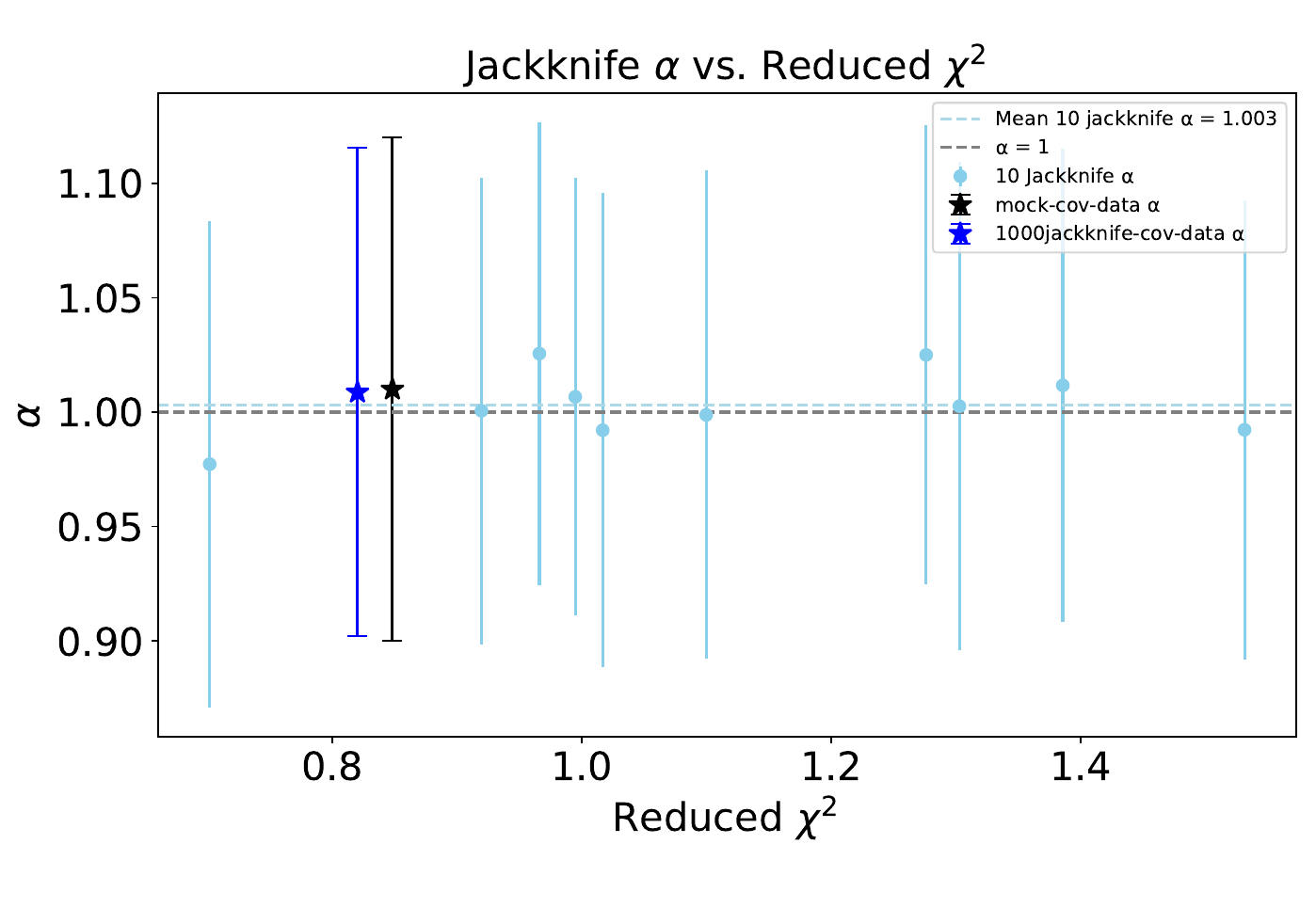}
    \caption{Jackknife robustness test performed by dividing the eBOSS and LoTSS masks into ten subregions and computing the eBOSS $\times$ LoTSS angular power spectrum for the first redshift bin, $0.6 < z < 0.66$, using a leave-one-out approach. It illustrates the 10 $\alpha$ measurements along with their associated 68\% confidence interval as a function of their reduced $\chi^2$ values (light blue points). The black pentagon represents the $\alpha$ measurement obtained from the real data measurement using covariance generated from \texttt{FLASK} mock, corresponding to Case 1 in Table~\ref{table:main_res}. The blue pentagon represents the result using 1000-subregion jackknife covariance with real data measurement.}
    \label{fig:jackknife}
\end{figure}

To assess the stability of our BAO measurement, we performed jackknife resampling tests. Figure~\ref{fig:jackknife} presents the results for both the 10-jackknife and 1000-jackknife analyses. The mean and dispersion of $\alpha$ derived from these tests are compared to the full dataset result. 

As discussed in the main text, the jackknife-derived error bars are slightly smaller than those from the real dataset due to reduced cosmic variance, while the mean values of $\alpha$ remain stable. The reduced chi-squared values further confirm the goodness of fit.

These figures complement our main findings and reinforce the reliability of our measurement.

\end{appendix}

\end{document}